\documentclass[pra, superscriptaddress, twocolumn]{revtex4}
\usepackage{optidef}

\usepackage[SquareTraceBrackets]{quantum}
\usepackage{textcomp}
\usepackage{graphicx,bm,natbib,upgreek,amsmath,mathrsfs,accents,easybmat}
\usepackage{amsbsy}
\usepackage{amsfonts}
\usepackage[dvipsnames]{xcolor}
\definecolor{myblue}{named}{MidnightBlue}

\usepackage[colorlinks = true,
            linkcolor = myblue,
            urlcolor  = myblue,
            citecolor = myblue,
            anchorcolor = myblue]{hyperref}

\usepackage{amsthm}

\newtheorem{theorem}{Theorem}

\newtheorem{proposition}[theorem]{Proposition}

\newtheorem{remark}{Remark} 

\usepackage{graphicx}

\usepackage{tabularx}
\DeclareGraphicsExtensions{.pdf, .jpg, .eps, .svg, .png}

\def\<{\langle}
\def\>{\rangle}

\newcommand{\SU}[1][2]{{\rm SU} (#1)}
\newcommand{\su}[1][2]{{\rm su} (#1)}
\newcommand{\SO}[1][3]{{\rm SO} (#1)}
\newcommand{\iu}{\mathrm{i}\mkern1mu}
\newcommand\scalemath[2]{\scalebox{#1}{\mbox{\ensuremath{\displaystyle #2}}}}
\newtheoremstyle{paragrapho}{}{}{}{\parindent}{}{~---~}{0em}{{\bfseries \thmnote{#3}}}
\theoremstyle{paragrapho}
\newtheorem{paragrapho}{}[]

\begin{document}

\title{Security of hybrid BB84 with heterodyne detection}

\author{Jasminder S. Sidhu}
\email{jsmdrsidhu@gmail.com}
\affiliation{SUPA Department of Physics, The University of Strathclyde, Glasgow, G4 0NG, UK}

\author{Rocco Maggi}
\affiliation{Dipartimento  Interateneo di Fisica, Politecnico di Bari, 70126, Bari, Italy}

\author{Saverio Pascazio}
\affiliation{Dipartimento  Interateneo di Fisica, Universit\`a di Bari, 70126, Bari, Italy}
\affiliation{INFN, Sezione di Bari, 70126 Bari, Italy}

\author{Cosmo Lupo}
\email{cosmo.lupo@poliba.it}
\affiliation{Dipartimento  Interateneo di Fisica, Politecnico di Bari, 70126, Bari, Italy}
\affiliation{Dipartimento  Interateneo di Fisica, Universit\`a di Bari, 70126, Bari, Italy}
\affiliation{INFN, Sezione di Bari, 70126 Bari, Italy}

\date{\today}

\begin{abstract}
Quantum key distribution (QKD) promises everlasting security based on the laws of physics. Most common protocols are grouped into two distinct categories based on the degrees of freedom used to carry information, which can be either discrete or continuous, each presenting unique advantages in either performance, feasibility for near-term implementation, and compatibility with existing telecommunications architectures. Recently, hybrid QKD protocols have been introduced to leverage advantages from both categories. In this work we provide a rigorous security proof for a protocol introduced by Qi in 2021, where information is encoded in discrete variables as in the widespread Bennett Brassard 1984 (BB84) protocol but decoded continuously via heterodyne detection. Security proofs for hybrid protocols inherit the same challenges associated with continuous-variable protocols due to unbounded dimensions. Here we successfully address these challenges by exploiting symmetry. Our approach enables truncation of the Hilbert space with precise control of the approximation errors and lead to a tight, semi-analytical expression for the asymptotic key rate under collective attacks. As concrete examples, we apply our theory to compute the key rates under passive attacks, linear loss, and Gaussian noise.
\end{abstract}

\maketitle

\section{Introduction}
\label{sec:intro}

\noindent
Quantum key distribution (QKD) exploits quantum optics to establish secret keys between distant users over an insecure communication channel~\cite{Pirandola2020_AOP}. Unlike software-based solutions such as the Rivest-Shamir-Adleman (RSA) protocol~\cite{Rivest1978_CACM} and post-quantum cryptography~\cite{Bernstein2009_book}, QKD promises informational-theoretical security under a well-defined set of assumptions~\cite{Mosca2013}. This means that keys obtained through QKD protocols bear the property of everlasting security; they remain secure against any future development in algorithms, supercomputers, and quantum computers~\cite{Mosca2010}. Most QKD protocols cluster into two categories: discrete-variable (DV) QKD and continuous-variable (CV) QKD, which differ in the degrees of freedom used to encode information. DV protocols, such as the celebrated BB84 (Bennett and Brassard, 1984)~\cite{Bennett1984} use discrete degrees of freedom, such as polarisation or time-bin coding and decode information through direct detection~\cite{Bennett1992_PRL, Ekert1991_PRL}. CV QKD protocols instead exploit the continuous amplitude and phase quadratures of the optical field to encode information, and coherent detection, such as homodyne and heterodyne detection, for decoding~\cite{Ralph1999_PRA, Grosshans2002_PRL, Weedbrook2012_RMP}. 

Both categories of QKD protocols have separate features that lend themselves to different applications~\cite{DiamantiLo}. DV protocols are generally more robust to channel losses, which permit their implementation in long-range ($\gtrsim$100~km) quantum communications such as satellite-based networks~\cite{Sidhu2021advances, sidhu2021key, Wallnofer2021, Islam2022finite}. This has been showcased through recent landmark satellite-based QKD experiments and quantum networking demonstrations~\cite{Lu2022_RMP}, with the current state of the art for entanglement distribution being over 1200~km~\cite{Yin2017_S}. However, DV protocols rely on high-efficiency photon detectors, which are currently costly and bulky owing to their need for cryogenics~\cite{Holzman2019_AQT}. CV QKD protocols find more widespread implementation in terrestrial networks ($\lesssim$100~km) given their compatibility with existing telecommunication infrastructures and tolerance to co-propagation with classical signals~\cite{Karinou2018_IEEE}. While the feasibility of CV protocols over greater ranges have been explored~\cite{Dequal2021_npjQI}, distances with CV protocols are typically shorter than with DV protocols since security requires very low noise during transmission and detection~\cite{Nitin2022}.

With DV and CV protocols offering distinct advantages, a hybrid protocol that merges the salient features of both may offer a promising route to enable longer-range quantum networking. Specifically, this hybrid protocol would operate with low-cost photodiodes at room temperature and improved compatibility with existing telecommunication architectures. Additionally, by benefiting from mature security proofs, hybrid protocols could contribute to advancing the state-of-the-art in networked quantum communications over global scales. Developing such hybrid protocols for QKD has recently gained traction, where information is encoded in discrete variables of light and decoded through coherent detection~\cite{Qi2021_PRA, Primaatmaja2022_Q}.

The asymptotic key rate, obtained after an asymptotic number of transmitted signals, is regularly used as a proxy for upper bounding the performance of QKD protocols and to provide straightforward comparison between protocols. A severe bottleneck is that security proofs require a theoretical model that precisely matches the physical devices used in an implementation. Hybrid protocols additionally inherit technical challenges associated with CV protocols due to an infinite-dimensional Hilbert space. There have been a number of approaches to overcome these challenges to simplify security analyses and simplify key rate calculations. First, by coarse-graining measurement outcomes from heterodyne detection, the discrete variables encoded in the input signals can be inferred at the expense of increased noise~\cite{Qi2021_PRA}. Heterodyne detection also enables full reconstruction of the photon number distribution of received signals, which can be exploited to bound the number of bits leaked to the environment~\cite{Qi:het, Chapman:het}. Second, lower bounds have been exploited to reduce the key-rate calculation to a semi-definite program~\cite{Primaatmaja2022_Q}, an approach also suitable to encompass decoy states. 

Hybrid QKD has been explored with single photons, employing either polarisation or time bin encoding~\cite{Qi2021_PRA}, decoy states~\cite{Primaatmaja2022_Q}, and discrete modulation phase-shift keying~\cite{Lin2019_PRX}. In this work, we explore the potential of single-photon-based hybrid QKD for practical implementation and deployment across quantum-secured networks. Specifically, we improve the security analysis within the collective attack framework to establish a tight lower bound on the asymptotic key rate. The tightness of our method enables higher key rates and increased robustness to noise over the previous single-photon based hybrid protocol. We quantify this improvement within an experimentally feasible parameter space, providing insights into the current readiness for implementation. Additionally, we compare the performance of hybrid protocols with DV and CV protocols to discuss their current viability for applications in quantum networking, which remains an open question in the field. Section~\ref{sec:results_summary} provides an executive summary of our results, with an outline of the paper provided in section~\ref{sec:outline}.


\subsection{Summary of results}
\label{sec:results_summary}

\noindent
In this work, we provide the first rigorous security proof that yields a tight lower bound on the key rate. We obtain a semi-analytical expression for the asymptotic key rate under collective attacks, where attacks from an eavesdropper on transmitted signals are identical and statistically independent. It is likely that methods developed in the context of of DV or CV QKD~\cite{PhysRevLett.114.070501, PhysRevLett.118.200501, Lin2019_PRX, Kanitschar2023_PRXQ, Lupo2022_PRXQ, AAcin} can be adapted and applied to hybrid protocols too.

To explore the performance of hybrid QKD, we outline a general approach to exploit state symmetries to establish invariant states with reduced parameterisation. Note that security proofs against general attacks often introduce a symmetrisation step to endow composite systems with permutation invariance. CV QKD protocols have been shown to additionally exhibit Lie group invariance. The symmetry we appeal to is the invariance of two-party composite states under $(U \otimes U^*)$ transformations, where $U$ belongs to the SU(2) Lie group that physically represents a linear-optics passive unitary acting on the two polarisation modes. The resulting invariant states we derive have a significantly reduced parameterisation; one that scales linearly with the Hilbert space dimension, compared with the quadratic scaling of the original composite states. This reduced parameterisation permits efficient numerical calculation of the secret key rate. 

Inspired by the numerical approach developed in Refs.~\cite{Coles2012, Coles2016_NC, Winick2018_Q}, we use our invariant states to construct a constrained key rate optimisation that is closely aligned to an experimental implementation of the protocol. In particular, we constrain the optimisation according to error parameters that can be directly measured, including the gain $Q$ and the quantum bit error rate (QBER). Our work therefore provides a route towards an experimental realisation of the hybrid QKD analysed in this work. Most notably, this procedure allows us to perform an exact numerical optimisation with full control of the  error due to finite-dimensional cut-off of the otherwise infinite-dimensional Hilbert space that characterises CV QKD systems. 

By exploring the performance of a pure loss channel, we find the asymptotic key rate for our hybrid protocol scale as O($\eta^2$), where $\eta$ is the attenuation factor. Most DV and CV protocols are characterised by a linear scaling O($\eta$). The worse scaling of hybrid protocols than discrete-variable ones is due to decreasing gain with increasing range, and is the penalty to pay for improved compatibility with terrestrial networks. This work is the first to quantify this tradeoff that would be instrumental in guiding future research into the use of hybrid protocols for quantum networking. 

For passive attacks, our theory provides higher rates and can tolerate higher channel losses than what estimated by previous security analyses. When Gaussian noise is introduced the key rate decreases rapidly, highlighting high sensitivity of hybrid protocols to excess noise in the detector; a feature inherited from continuous variable protocols. For an excess noise variance of $N = 10^{-4}$ (in shot-noise units), we demonstrate our hybrid scheme can tolerate losses up to $\sim$17~dB, making it suitable to deliver high-rate QKD in terrestrial or free-space quantum networks over metropolitan scales. However, the key rates are lower than those achieved with CV protocols. This suggests that the hybrid approach is not always advantageous in terms of robustness to noise.

Before concluding, we note that our hybrid QKD protocol significantly eases implementation over DV and CV protocols. First, in contrast to DV QKD, our hybrid protocol allows for the use of faster receivers and does not require sifting since a single decoding measurement applies to both encoding bases. Second, in contrast to CV QKD, our hybrid protocol does not require a shared local oscillator or a pilot tone. This significantly reduces transmitter and receiver complexity and the potential for side-channel attacks~\cite{Ma2013_PRA, Jouguet2013_PRA}. Combined with a key rate optimiser that is closely aligned to an experimental implementation, our work provides a feasible route towards practical implementation of the protocol.


\subsection{Outline of paper}
\label{sec:outline}

\noindent
The paper develops as follows. In Section~\ref{sec:id} we review the protocol introduced in Ref.~\cite{Qi2021_PRA} based on independent detection of two polarisation modes. In Section \ref{sec:security} we lay the foundation of our security analysis, which is inspired by the work in Refs.~\cite{Coles2012, Coles2016_NC, Winick2018_Q}. A first case study is presented in Section~\ref{sec:pure_loss}, which explores a pure-loss communication channel. Section~\ref{sec:symmetries} extends our approach to most general collective attacks. Here we introduce symmetry in the protocol and exploit it to simplify the security analysis. In Section~\ref{sec:minimisation_rel_ent} we show how symmetry allows us to control the error introduced by the truncation of the Hilbert space in view of the numerical optimisation. This approach is developed in Section~\ref{sec:Gauss} to study in detail the case of Gaussian noise. This noise model may be used to describe electronic noise in heterodyne detection. Conclusions and discussions are presented in Section~\ref{sec:end}, where we summarise the motivations for our work and the most important take-home messages. Further details and a number of technical results are reported in the Appendix.


\section{BB84 with heterodyne detection - independent detection}
\label{sec:id}

\noindent
The subject of our analysis is one of the hybrid protocols introduced by Qi in Ref.~\cite{Qi2021_PRA}; one that is based on \textit{independent detection} of the two optical modes used to encode quantum information. To make the presentation more concrete, we assume that these modes represent polarisation, though other degrees of freedom could be equivalently considered.

We first start with a review of the hybrid protocol in the prepare-and-measure (PM) representation. The schematic setup of our protocol is shown in Fig.~\ref{fig:schematic} and the protocol is as follows:

\begin{enumerate}

\item \textit{State preparation.} First, as in BB84, the sender (Alice) encodes one bit of information by preparing a single-photon state with either horizontal ($H$) or vertical ($V$) polarisation. Alternatively, Alice may use diagonal ($D$) or anti-diagonal ($A$) polarisation. The choice of polarisation basis is random; without loss of generality we assume equal probabilities. For each transmission round, Alice sends her prepared states to Bob through an insecure quantum channel.

\item \textit{Measurement.} The hybrid protocol differs from standard BB84 in the measurement procedure. In standard BB84 the receiver (Bob) applies photon-detection to decode received signals. Instead, we assume coherent decoding by heterodyne detection, which is a CV measurement defined on a single mode of the quantum electromagnetic field~\cite{Paris_book}. Bob receives two optical modes, characterised by the canonical bosonic annihilation and creation operators $\{ b_H, b_H^\dagger \}$ and $\{ b_V, b_V^\dagger \}$, which corresponds to $H/V$ polarisation. The canonical operators for the $D/A$ polarisation are obtained from the latter as 
\begin{align}
b_D & = ( b_H + b_V)/\sqrt{2} \, , \label{oper1} \\
b_A & = ( b_H - b_V)/\sqrt{2} \, . \label{oper2}
\end{align}
Bob performs a heterodyne measurement on the state, which is described by a continuous family of positive operator-valued measurement (POVM) elements
\begin{align}
\Lambda(\beta) = \frac{1}{\pi} \, |\beta \rangle \langle \beta| \, ,
\end{align} 
where $|\beta \rangle$ is the coherent state of amplitude $\beta = ( q + i p )/\sqrt{2}$ of the optical mode being measured. Recall that the coherent states satisfy the completeness relation, from which we obtain
\begin{align}
\int d^2\beta \Lambda(\beta) = I \, ,
\end{align}
where $I$ is the identity operator and $d^2\beta := dq dp/2$.

Alice and Bob repeat the state preparation and measurement stage $m$ times.

\item \textit{Basis announcement.} Alice announces her choices for the polarisation basis. According to this information Bob will adapt his inference strategy. However, as remarked below, no sifting is necessary.

\begin{figure}[t!]
\centering
\includegraphics[width=1\columnwidth]{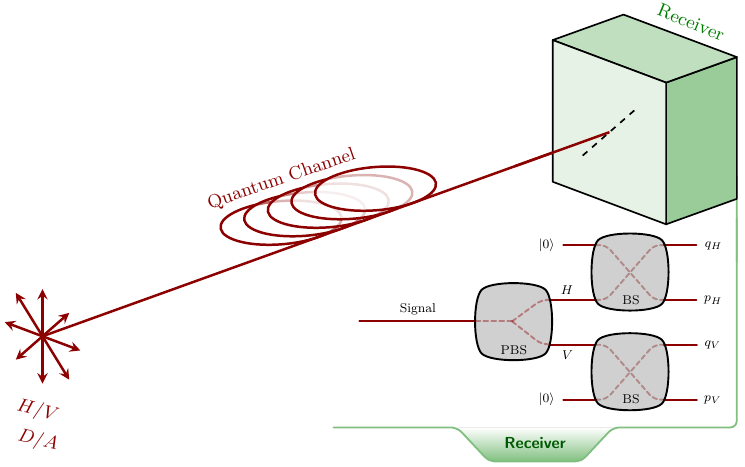} 
\caption{
\textbf{Schematic setup of the protocol}: The transmitter (Alice) prepares polarisation states in the rectilinear ($H/V$) or diagonal ($D/A$) bases. After the quantum channel, the receiver (Bob) performs a heterodyne measurement on received states. The inset illustrates the receiver in detail:  a polarising beam splitter (PBS) sorts the beam according to polarisation, then each beam is measured using heterodyne detection. }
\label{fig:schematic}
\end{figure}%

\item \textit{Inference.} To infer the bit value encoded by Alice, Bob compares the output of mode-wise heterodyne to a given threshold value $\tau >0$ that is decided before executing the protocol. Consider the operators 
\begin{align}
R_0 & = \int_{|\beta|^2\leq \tau} d^2 \beta \, \Lambda(\beta) \, , \\
R_1 & = \int_{|\beta|^2 > \tau} d^2 \beta \, \Lambda(\beta) \, .
\end{align}
Bob then establishes a key map through a threshold detection obtained by combining these operators on the two modes. For example, to discriminate between $H$ and $V$ polarisation, we need to combine the above operators applied to each mode of polarisation, denoted as $R_0^H$, $R_1^H$ and $R_0^V$, $R_1^V$. The threshold detection corresponds to the POVM elements
\begin{align}
M_H & = R_1^H \otimes R_0^V \, , \label{MH} \\
M_V & = R_0^H \otimes R_1^V \, . \label{MV}
\end{align}
To obtain a complete set, one also needs to introduce the null operator
\begin{align}
M_0 = I - M_H - M_V \, ,
\end{align}
in such a way that $M_H + M_V + M_0 = I$. Successful detection is associated to measurement outcomes $M_H$ (in which case Bob infers a horizontally polarised photon) or $M_V$ (Bob infers vertical polarisation). Events corresponding to the null outcome $M_0$ are discarded. The analogous construction applied to $D/A$ polarisation leads to the definition of the operator $M_D$, $M_A$.

\end{enumerate}

The above describes the quantum part of the QKD protocol. Alice and Bob use the data collected to determine the secret key rate by solving the optimisation problem in Eq.~\eqref{eqn:min_entropy}. The protocol is aborted if no secret keys can be generated, otherwise, they proceed. The raw keys are finally post-processed for parameter estimation, error correction, and privacy amplification. The post-processing procedures are equivalent to standard BB84.

\begin{remark}
One interesting feature of this hybrid protocol is that Bob only needs to apply heterodyne detection to infer both the $H/V$ and $D/A$ modes of polarisation. This is because from Eqs.~(\ref{oper1})-(\ref{oper2}) the outcomes $\beta_D$, $\beta_A$ of heterodyne detection in the $D/A$ polarisation modes can be obtained exactly from the outcomes $\beta_H$, $\beta_V$ of heterodyne detection in the $H/V$ modes,
\begin{align}
\beta_D & = ( \beta_H + \beta_V)/\sqrt{2} \, , \label{het1} \\
\beta_A & = ( \beta_H - \beta_V)/\sqrt{2} \, . \label{het2}
\end{align}
This implies that no data is discarded, in contrast to the sifting phase in standard BB84. The price to pay, as highlighted in Ref.~\cite{Qi2021_PRA}, is an additional error in the inference compared to direct detection.
\end{remark}

\begin{figure}[t!]
\centering
\includegraphics[width=0.95\columnwidth]{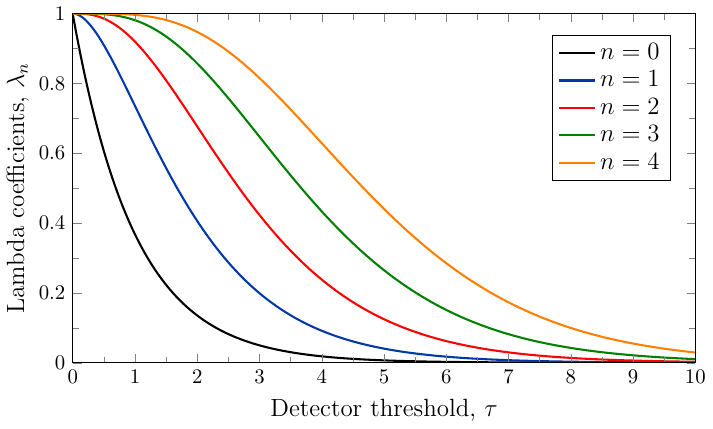} 
\caption{\textbf{Eigenvalues of $\bm{R_1}$ operator}: The coefficients $\lambda_n$ in Eq.~(\ref{eq:lambdas}) plotted vs the threshold value $\tau$, from bottom to top, $n=0,1,2,3,4$.}
\label{fig:lambdas}
\end{figure}

We conclude this section by presenting the expansion of the operators $R_0$, $R_1$ in the number basis. From the expression of the coherent state, 
$|\beta \rangle = e^{-|\beta|^2/2} \sum_{n=0}^\infty \beta^n / \sqrt{n!} | n\rangle$, we obtain
\begin{align}
R_0 & = \sum_{n=0}^\infty  (1-\lambda_n) \, | n \rangle \langle n |  \, , \label{R0def} \\
R_1 & = \sum_{n=0}^\infty  \lambda_n \, | n \rangle \langle n |  \, ,
        \label{R1def}
\end{align}
where
\begin{align}\label{eq:lambdas}
    \lambda_n : = \frac{\Gamma[1+n,\tau]}{n!} 
\end{align}
and $\Gamma$ is the incomplete gamma function. Figure~\ref{fig:lambdas} shows a plot of these coefficients versus the threshold value $\tau$.

%


\section{Security analysis}\label{sec:security}

\noindent
The security of our hybrid protocol is better assessed using its equivalent entanglement-based (EB) representation. We emphasise that the EB representation provides a useful mathematical tool but the physical implementation of the protocol in general follows the PM representation.

In the EB representation Alice prepares a pair of photons that are entangled in polarisation,
\begin{align}\label{inputstate}
|\phi\rangle_{AA'} & = ( |H\rangle_A |H\rangle_{A'} + |V\rangle_A |V\rangle_{A'} ) /\sqrt{2} \, .
\end{align}
Alice sends photon $A'$ to Bob and keeps photon $A$ for herself. Alice eventually measures photon $A'$ in either the $H/V$ basis or in the conjugate $D/A$ basis, thus conditionally preparing the other photon in the same state of polarisation.

A noisy communication channel $\mathcal{N}_{A' \to B}$ maps the state $|\phi\rangle_{AA'}$ into
\begin{align}\label{rhoAB}
    \rho_{AB} = I_A \otimes \mathcal{N}_{A' \to B} 
    ( |\phi\rangle \langle \phi| ) \, ,
\end{align}
where $I_A$ is the identity channel acting on photon $A$. In this work, we consider collective attacks, where the eavesdropper applies i.i.d.~noise to each signal transmission. Hence, for $m$ photons sent by Alice, the state shared with Bob is simply given by the tensor power $\rho_{AB}^{\otimes m}$. In the limit of $m \to \infty$, the asymptotic secret key rate rate, expressed in secret bits per photon sent can be expressed as~\cite{Coles2012, Coles2016_NC, Winick2018_Q} 
\begin{align}
    r(\rho_{AB}) = D\left[ 
    \mathcal{G}(\rho_{AB}) \| \mathcal{Z}(\mathcal{G}(\rho_{AB})) 
    \right] 
    - \mathrm{leak_{EC}} \, ,
    \label{eqn:DW}
\end{align}
where $D\left[  \rho  \| \sigma  \right] = \mathrm{Tr} ( \rho \log{\rho} ) - \mathrm{Tr} ( \rho \log{\sigma} )$ is the quantum relative entropy ($\log$ denotes the logarithm in base $2$, whereas $\ln$ is used for natural logarithm), and the maps $\mathcal{G}$ and $\mathcal{Z}$ will be defined below. The relative entropy term quantifies the number of secret bits per photons that can be extracted from the raw key after privacy amplification. The final secret key rate is then determined by subtracting the term $\mathrm{leak_{EC}}$, which is the number of bits per photon leaked for error correction. Here we assume reverse reconciliation.

The map $\mathcal{G}$ in Eq.~\eqref{eqn:DW}, dubbed \textit{key map}, is a partial isometry that gives a coherent representation of the measurement and decoding applied by the receiver. It takes as input a state of the $B$ system and outputs a state of the composite system $B B_1$, where $B_1$ is an auxiliary qubit: 
\begin{align}
\mathcal{G}(\rho_{AB}) = ( I \otimes K ) \rho_{AB} ( I \otimes K^\dagger ) \, ,
\end{align}
where
\begin{align}
K =  |H\rangle_{B_1} \otimes \sqrt{ M_H } + |V\rangle_{B_1} \otimes \sqrt{ M_V } \, ,
\label{eqn:map}
\end{align}
with $M_H$ and $M_V$ as in Eqs.~(\ref{MH})-(\ref{MV}). The state $\mathcal{G}(\rho_{AB})$ is in general not normalised. Its trace determines the gain $Q$ such that
\begin{align} 
Q = \mathrm{Tr}[\mathcal{G}(\rho_{AB})] & = \mathrm{Tr} [( I \otimes K^\dagger K ) \rho_{AB} ] \\
& = \mathrm{Tr} [( M_H + M_V )\rho_{B} ] \, .
\label{P0def}
\end{align}
The gain is the probability that Bob obtains a valid measurement output and can be estimated in an experimental implementation of the protocol.

The map $\mathcal{Z}$ in Eq.~\eqref{eqn:DW} applies the \textit{pinching map} to the auxiliary system $B_1$, inducing complete dephasing in the basis $\{ |H\rangle_{B_1}, |V\rangle_{B_1} \}$,
\begin{align}\label{Zmapdef}
\mathcal{Z}( \mathcal{G}(\rho_{AB}) )  & =  |H\rangle_{B_1} \langle H| \mathcal{G}(\rho_{AB}) |H\rangle_{B_1} \langle H| \nonumber \\
& \phantom{=}~ + | V \rangle_{B_1} \langle V | \mathcal{G}(\rho_{AB}) | V \rangle_{B_1} \langle V|  \, .
\end{align}
An analogous definition may be introduced for the $D/A$ polarisation modes. However, as in our discussion we will only consider symmetric states, it is sufficient to consider the $H/V$ basis.

The calculation of the relative entropy is simplified using proposition:
\begin{proposition}
\label{prop:rel_ent}
The relative entropy equals the difference of two entropies:
\begin{align}
    D\left[ \mathcal{G}(\rho_{AB}) \| \mathcal{Z}(\mathcal{G}(\rho_{AB})) \right] = S[\mathcal{Z}(\mathcal{G}(\rho_{AB}))] - S[\mathcal{G}(\rho_{AB})] \, ,
\end{align}
where $S[\sigma] = - \mathrm{Tr}( \sigma \log{\sigma})$ is the von Neumann entropy.
\end{proposition}

\noindent
\begin{proof}
Note that Eq.~(\ref{Zmapdef}) implies
\begin{align}
    \log{\left[ \mathcal{Z}(\mathcal{G}(\rho_{AB})) \right] } & = | H \rangle_{B_1} \langle H |
    \log{ \left[  \mathcal{Z}(\mathcal{G}(\rho_{AB}))  \right] }| H \rangle_{B_1} \langle H | \nonumber \\
& \phantom{=}~ + | V \rangle_{B_1} \langle V | \log{  \left[ \mathcal{Z}(\mathcal{G}(\rho_{AB}))  \right]  } | V \rangle_{B_1} \langle V | \, .
\end{align}

Therefore
\begin{align}
&     \mathrm{Tr} \left\{ \mathcal{G}(\rho_{AB})  \log{ \left[  \mathcal{Z}(\mathcal{G}(\rho_{AB}))  \right]  } \right\} \nonumber \\
& = \mathrm{Tr} 
\left\{ \left( | H \rangle_{B_1} \langle H | \mathcal{G}(\rho_{AB}) | H \rangle_{B_1} \langle H |  \right. \right. \nonumber \\
& \phantom{===}~ + \left. \left. | V \rangle_{B_1} \langle V | \mathcal{G}(\rho_{AB}) | V \rangle_{B_1} \langle V |  \right)  \log{ 
    \left[ \mathcal{Z}(\mathcal{G}(\rho_{AB})) \right] }  \right\} \\
& = \mathrm{Tr}  \left\{ \mathcal{Z}(\mathcal{G}(\rho_{AB})) \log{ \left[  \mathcal{Z}(\mathcal{G}(\rho_{AB}))  \right] } \right\} \, .
\end{align}
\end{proof}

To simplify the notation for the rest of the paper, we denote
\begin{align}
D[ \rho_{AB} ] :=  D\left[ \mathcal{G}( \rho_{AB} ) \| \mathcal{Z}(\mathcal{G}(\rho_{AB} )) \right] \, .
\end{align}

Besides the gain $Q$, another parameter that can be estimated experimentally is the qubit error rate (QBER) $E$. First consider the quantity
\begin{align}\label{c_def}
    c := 
    \frac{1}{2} \mathrm{Tr} \left[ 
    ( 
    |H\rangle \langle H | \otimes M_V 
    +
    |V\rangle \langle V | \otimes M_H
    )
    \rho_{AB} 
    \right] 
    \, .
\end{align}

From $c$ and $Q$ we obtain the QBER
\begin{align}
E = \frac{2c}{Q}
= \frac{
    \mathrm{Tr} \left[ 
    ( |H\rangle \langle H | \otimes M_V 
+
|V\rangle \langle V | \otimes M_H     )
    \rho_{AB} 
    \right] 
}{ \mathrm{Tr} [
( M_H + M_V )
\rho_{B} ] } 
 \, ,
\end{align}
such that $Q E = 2c$. In turn, from the QBER we estimate the error correction term in the key rate,
\begin{align}\label{leakEC}
    \mathrm{leak_{EC}} 
    = Q h_2( E ) \, ,
\end{align}
where $h_2(x) = - x \log{x} - (1-x) \log{(1-x)}$ is the binary Shannon entropy. This expression follows from the model of symmetric binary channel~\cite{Cover2006book}.

Finally, we remark that in QKD we do not assume complete knowledge of the state $\rho_{AB}$, therefore one should consider the worst-case scenario that is compatible with the experimental data. In our setup, the experimental data allows Alice and Bob to estimate the parameters $Q$ and $c$. Furthermore, as discussed in~\cite{Qi:het}, heterodyne detection allows Bob to estimate the photon-number distribution of the unknown state $\rho_{AB}$,
\begin{align}
P_j :=  \sum_{a = 0}^j \mathrm{Tr}  \left[ \left(  |a\rangle_H \langle a| + |j - a\rangle_V \langle j - a|  \right)  \rho_{B} \right]  \, .
\end{align}

In conclusion, the asymptotic key rate is obtained by solving the following constrained minimisation problem
\begin{align}\label{eqn:min_entropy}
    \min_{\rho_{AB} \in \mathcal{S}} D[ \rho_{AB} ]  - Q \, h_2( E ) \, ,
\end{align}
given experimental estimates for $Q$ and $E$, and the set $\mathcal{S}$ of feasible states defined through the following conditions:
\begin{enumerate}
\item The reduced state of Alice photon is maximally mixed:
\begin{align}
\rho_A =  \mathrm{Tr}_B (\rho_{AB}) = I/2  = \frac{|H\rangle \langle H| + |V\rangle \langle V|}{2}\, .
\end{align}

\item The experimentally estimated error parameter $c$. In full generality, one should distinguish between errors in the $H/V$ basis and those in the $D/A$ basis. Here, for simplicity we assume that they are independent of the polarisation direction. We will use and expand this symmetry assumption below.
\begin{align}
    & \mathrm{Tr}[ ( |H\rangle \langle H | \otimes M_V ) \rho_{AB}] =
    \mathrm{Tr}[ ( |V\rangle \langle V | \otimes M_H ) \rho_{AB} ] \nonumber\\
    & = \mathrm{Tr}[ ( |D\rangle \langle D | \otimes M_A ) \rho_{AB} ] =
    \mathrm{Tr}[ ( |A\rangle \langle A | \otimes M_D ) \rho_{AB} ] \nonumber \\
    & = c \, .
\end{align}

\item The experimental estimated gain $Q$. As for $c$, we should make a distinction between the two polarisation bases. For simplicity we assume equal value for both.
\begin{align} 
\mathrm{Tr} [ (M_H + M_V ) \rho_{B} ]  = \mathrm{Tr} [ (M_D + M_A ) \rho_{B} ]  & = Q \, .
\label{P0def}
\end{align}

\item The experimental estimates $P_j$ for the photon number distribution, up to a certain photon number $k$. For $j = 0, \dots, k$:
\begin{align}
\sum_{a = 0}^j \mathrm{Tr}  \left[ \left(  |a\rangle_H \langle a| + |j - a\rangle_V \langle j - a|  \right)  \rho_{B} \right]  = P_j  \, .
\end{align}

\end{enumerate}
Note the final three constraints originate from experimental informed parameters, providing a route to physical implementation of the protocol. In addition, the asymptotic key rate in Eq.~\eqref{eqn:min_entropy} can be further maximised by optimising over the detector threshold $\tau>0$.


\section{Pure-loss channel}\label{sec:pure_loss}

\noindent
As a first example, here we determine the asymptotic secret key rate for a pure-loss channel. The communication channel $\mathcal{N}$ is a wiretap channel that induces polarisation-independent loss with transmissivity factor $\eta \in [0,1]$. In the Heisenberg picture, the canonical operators are transformed as follows
\begin{align}
b_H & \to \sqrt{\eta} \, b_H + \sqrt{1-\eta} \, e_H \, , \\
b_V & \to \sqrt{\eta}\,  b_V + \sqrt{1-\eta} \, e_V \, ,
\end{align}
where $e_H$, $e_V$ are auxiliary vacuum modes.

In the Schr{\"o}edinger picture, the input state (\ref{inputstate}) is transformed according to Eq.~(\ref{rhoAB}) into
\begin{align}
    \rho_{AB}  & = \eta |\phi \rangle_{AB} \langle \phi| + \frac{(1-\eta)}{2} I_A \otimes |0\rangle_B \langle 0| \, .
\label{eqn:sigma}    
\end{align}
where $|0\rangle_B$ is the vacuum state on Bob's side. From this expression we compute
\begin{align}
    D [ \rho_{AB} ] & =  2(1-\eta)(1-\lambda_0)\lambda_0  +  \eta (\lambda_0 + \lambda_1 - 2 \lambda_0 \lambda_1)  \, ,  \label{Poloss0988}  \\
    c &= \frac{\lambda_0}{2}  \left[  1  - ( 1 - \eta  )\lambda_0   - \eta \lambda_1  \right] \label{celoss0988} \, .
\end{align}
As expected for a pure-loss channel, all qubits that reach to Bob are secure, therefore we also obtain $Q= D [ \rho_{AB} ]$. 

\begin{figure*}
    \centering
    \includegraphics[width=\linewidth]{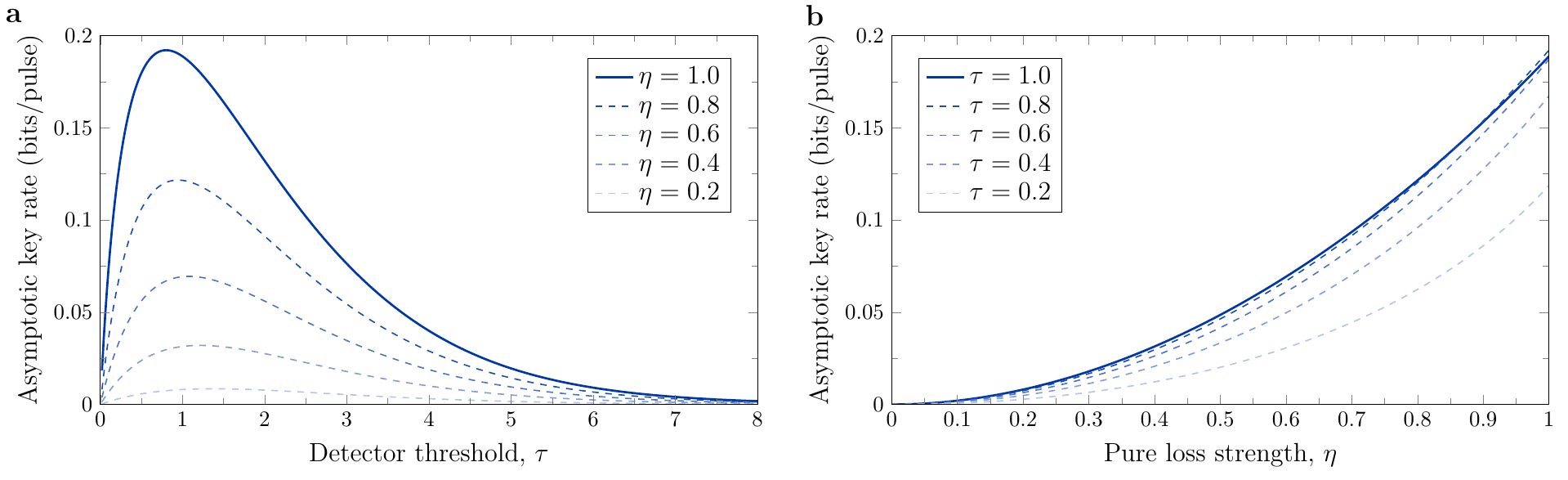}
    \caption{\textbf{Asymptotic key rate for pure loss}: a) Asymptotic rate as a function of the detector threshold $\tau$ for independent detection scheme, computed from Eq.~(\ref{eqn:asymrate_passive}). Solid line corresponds to lossless transmission, dashed to loss case with strength $\eta$, from bottom to top $\eta=0.2,0.4,0.6,0.8$. The optimal values for the detector threshold values that maximise the key rate depends on the noise: For transitivity $\eta=1$, $\tau_\text{opt}=0.8012$, $\eta=0.8$, $\tau_\text{opt}=0.9458$, $\eta=0.6$, $\tau_\text{opt}=1.0779$, $\eta=0.4$, $\tau_\text{opt}=1.2159$, and $\eta=0.2$, $\tau_\text{opt}=1.3768$. b) Asymptotic rate as a function of the pure loss strength $\eta$, computed for different values of $\tau$. } 
    \label{fig:pureloss_independent_detection}
\end{figure*}

Finally, the asymptotic secret key rate is obtained using Eqs.~(\ref{eqn:DW}) and (\ref{leakEC}):
\begin{align}\label{eqn:asymrate_passive}    
    r  = Q  \left( 1 - h_2\left( E \right)  \right) \, ,
\end{align}
with
\begin{align}
E = \frac{2c}{Q} = \frac{\lambda_0 [1 - (1 - \eta) \lambda_0 - \eta \lambda_1]}{ 2(1-\eta) \lambda_0 (1 - \lambda_0) + \eta (\lambda_0 + \lambda_1 - 2\lambda_0 \lambda_1) }  \, .
\end{align}
The key rate is illustrated in Fig.~\ref{fig:pureloss_independent_detection}(a) as a function of the threshold value $\tau$ for various transmissivity $\eta$. Note that the optimal detector threshold has a weak dependency on $\eta$. These optimal values are summarised in the caption to Fig.~\ref{fig:pureloss_independent_detection}. In Fig.~\ref{fig:pureloss_independent_detection}(b) we illustrate the dependence of the key rate on $\eta$, taking different values for $\tau$.

It is interesting to investigate the limit of large communication distance, i.e.~when $\eta \ll 1$. We obtain
\begin{align}
    r  & \simeq  \frac{ \eta^2 (\lambda_0 - \lambda_1)^2 }{4\lambda_0 (1-\lambda_0) \ln{2}}  = \frac{ \eta^2 \tau^2 }{( e^\tau -1 ) \ln{16}} \, .
\label{eqn:scaling}    
\end{align}
This shows the key rate is of order $O(\eta^2)$. The ultimate repeaterless PLOB (Pirandola-Laurenza-Ottaviani-Banchi) bound scales as $O(\eta)$~\cite{PLOB}. The same $O(\eta)$ scaling is commonly achieved by both DV and CV protocols. The sub-optimal scaling of the hybrid protocol is attributed to a decrease in the rate as the distance increases, which is driven by two independent mechanisms. First, it becomes increasingly unlikely that Bob receives transmitted photons leading to a decrease in the gain $Q$. Second, the QBER $E$ increases with the distance even for a pure-loss channel. Use of hybrid protocols must therefore address the tradeoff between sub-optimal scaling and increased compatibility. Finally, from Eq.~(\ref{eqn:scaling}) we obtain that in the limit of long distance the optimal value for the threshold is $\tau \simeq 1.59$.


\section{General collective attacks: exploiting symmetry}
\label{sec:symmetries}

\noindent
To assess the security of the hybrid protocol beyond the pure-loss channel, we must address two challenges:
\begin{enumerate}
    \item The Hilbert space associated with the receiver is infinite-dimensional. To implement the optimisation in Eq.~(\ref{eqn:min_entropy}) numerically, a cutoff into a finite-dimensional subspace is required, together with a method to control the cutoff error introduced.

    \item For a Hilbert space cutoff of up to $k$ photons on Bob side, the joint state $\rho_{AB}$ would lives in a space of dimensions $(k+1) (k+2)$~\footnote{There are $j+1$ ways to distribute $j$ photons on two polarisation modes. Therefore, the dimensions of Bob's Hilbert space are $\sum_{j=0}^k (j+1) = \frac{1}{2} (k+1) (k+2)$. Finally we need to multiply by $2$, which is the dimension of Alice's Hilbert space}. 
    This quadratic scaling with $k$ presents a bottleneck for efficient numerical optimisation. 
\end{enumerate}

Here we solve both issues by exploiting symmetry. Symmetry allows us to reduce the number of free parameters from quadratic to linear in the cutoff photon number $k$, also providing a way to control the error introduced by the Hilbert-space truncation.

The symmetry group is of the form $U \otimes U^*$, where the unitary $U$ is applied on Alice's system, and $U^*$ on Bob's.  Here $U$ denotes a linear-optics passive (LOP) unitary~\cite{Aniello2006} acting on the two polarisation modes, and $U^*$ is its complex-conjugate. Note that the Bell state in Eq.~(\ref{inputstate}) is invariant under $U \otimes U^*$ transformations.

LOP unitaries are defined as unitary transformations that, in the Heisenberg picture, act linearly on the canonical bosonic operators, without mixing the creation and the annihilation operators. Consider for example Alice's system, which is associated to the bosonic operators $\{ a_H, a_H^\dag \}$, $\{ a_V, a_V^\dag \}$ for horizontal and vertical polarisation respectively. A LOP unitary transforms the operators as follows
\begin{align}
    a_H & \to U a_H U^\dag = \alpha \, a_H + \beta \, a_V \, , \\
    a_V & \to U a_V U^\dag = -\beta^* \, a_H + \alpha^* \, a_V \, ,
\end{align}
where $\alpha$, $\beta$ are complex number such that $|\alpha|^2 + |\beta^2| = 1$. On Bob' side, the application of $U^*$ yields
\begin{align}
    b_H & \to U^* b_H (U^*)^\dag = \alpha^* \, b_H + \beta^* \, b_V \, , \\
    b_V & \to U^* b_V (U^*)^\dag = -\beta \, b_H + \alpha \, b_V \, .
\end{align}
We remark that LOP unitaries preserve the total photon number, i.e.,
\begin{align}
U ( a_H^\dag a_H + a_V^\dag a_V ) U^\dag & = a_H^\dag a_H + a_V^\dag a_V \, , \\
U^* ( b_H^\dag b_H + b_V^\dag b_V ) U^\mathsf{T} & = b_H^\dag b_H + b_V^\dag b_V \, .
\end{align}
This implies that states that are invariant under the action of the symmetry group $U \otimes U^*$ are block-diagonal in the total photon number both on Alice and on Bob side.

Note that in our protocol there is only one photon on Alice's side, whereas for a generic attack there could be an arbitrary distribution of photon number on Bob side. This leads to the following form for a state that is invariant under the symmetry group:
\begin{align}\label{symstate0}
    \rho_{AB}^{\text{(inv)}} = \sum_{j=0}^\infty P_j \rho_{1:j}^{\text{(inv)}} \, ,
\end{align}
where $\rho_{1:j}^{\text{(inv)}}$ is an invariant state with one photon on Alice side and $j$ photons on Bob side, and $P_j$ is the probability of having $j$ photons on Bob side.

In Appendix \ref{App:groupt} we derive explicit expressions for the invariant states $\rho_{1:j}^{\text{(inv)}}$. We show that for each $j>0$ there exists a one-parameter family of invariant states, $\rho_{1:j}^{\text{(inv)}}(f_j)$, with $f_j \in [0,1]$, whereas for $j=0$ the invariant state is unique. Note that for any $j \neq j'$
the states $\mathcal{G}( \rho_{1:j}^{\text{(inv)}} )$ and $\mathcal{G}( \rho_{1:j'}^{\text{(inv)}} )$ have orthogonal support, as well as $\mathcal{Z}(\mathcal{G}( \rho_{1:j}^{\text{(inv)}} ))$ and $\mathcal{Z}(\mathcal{G}( \rho_{1:j'}^{\text{(inv)}} ))$. This implies that the relative entropy in Eq.~(\ref{eqn:DW}) reads
\begin{align}\label{Dinv}
    D [ 
    \rho_{AB}^{\text{(inv)}}
    ] 
    =
    P_0 
    D [ 
    \rho_{1:0}^{\text{(inv)}}
    ]
    +
    \sum_{j=1}^\infty
    P_j   
    D [ 
    \rho_{1:j}^{\text{(inv)}} (f_j)
    ]
    \, .
\end{align}

For each $j$ we can define the corresponding parameter $c_j(f_j)$. By linearity, we have 
\begin{align}\label{cinv}
 c = P_0 c_0 + \sum_{j=1}^\infty  P_j c_{j}(f_j) \, .   
\end{align}
An analogous decomposition holds for the gain $Q$, i.e.,
\begin{align}
Q = \sum_{j=0}^\infty Q_{j} \, ,
\end{align}
where $Q_{j}$ is the gain subject to Bob receiving exactly $j$ photons. Following Ref.~\cite{Qi2021_PRA}, for each $j$ we write 
\begin{align}
 Q_j = P_j Y_j \, ,   
\end{align}
where $Y_j$ is the yield for given $j$. The feasible range and expressions for $Y_j$, $c_j$ are computed explicitly in Appendix~\ref{App:ranges} and Appendix~\ref{App:inv_states},
where we also note that $Y_{j}$ does not depend on $f_j$. By combining these parameters we obtain the QBER conditioned on Bob receiving $j$ photons, 
\begin{align}
E_j = \frac{2c_j}{Y_j} \, ,
\end{align}
such that
\begin{align}
E = \frac{\sum_j Q_j E_j}{ Q } \, .
\end{align}


\subsection{A modified protocol}
\label{sec:mod_protocol}

\noindent
Symmetry is commonly exploited to assess the security of QKD protocols. Examples are found in the literature for both DV~\cite{KrausPRL, KrausPRA, Renner2007, postsel} and CV~\cite{PhysRevLett.114.070501, LevGDeF, PhysRevLett.118.200501, Ghorai_sym, Kaur} systems.

To justify our use of the $U \otimes U^*$ symmetry, we introduce a modified protocol that includes an \textit{active symmetrisation} step. First we note that the hybrid BB84 protocol requires Alice and Bob to share a reference frame in order to agree on the orientation of the $H/V$ and $D/A$ polarisation states. In the original protocol it is implicitly assumed that this reference frame is fixed. To make the protocol explicitly invariant under $U \otimes U^*$ symmetry we need to modify it in such a way that Alice and Bob randomly change the shared reference frame at each photon transmission. In the EB representation, this invariance is equivalent to applying a random local LOP transformation of the form $U \otimes U^*$, mapping any joint state $\rho_{AB}$ into an invariant state,
\begin{align}
    \rho_{AB} \to 
    \rho_{AB}^{\text{(inv)}} = 
    \int d\mu_U (U \otimes U^*) \rho_{AB} (U \otimes U^*)^\dag \, ,
\end{align}
where $d\mu_U$ is the Haar measure on the group. From this observation we argue that, for the modified protocol, it is sufficient to consider invariant states in the minimisation of the relative entropy. The advantage is that the invariant states are block-diagonal in the number basis and can be decomposed as in Eq.~(\ref{symstate0}).

Note that in fact only Alice needs to physically apply the unitary $U$. For Bob it is sufficient to always apply the same measurements and simply modify the inference strategy according to the unitary $U$ (and to the basis choice communicated by Alice). It may be possible to prove that invariant states are optimal even without introducing the active symmetrisation on Alice side, in a way similar to Ref.~\cite{Ghorai_sym}. However, we do not address this question here leaving it for future work.


\subsection{Passive attacks}

\noindent
In this section, we apply our modified protocol to passive attacks, where the eavesdropper does not add photons into the channel. This limits the minimisation of the relative entropy to the vacuum and the sector of the Hilbert space with one photon:
\begin{align}
    \rho_{AB}^\text{(inv)} = (1-\eta) \rho_{1:0}^\text{(inv)} + \eta \rho_{1:1}^\text{(inv)} (f_1) \, ,
\end{align}
where $\eta$ is the channel transmissivity.
The explicit form of the states $\rho_{1:j}^\text{(inv)}$ and of the parameters $Y_j$, $c_{1:j}$ are presented in Appendix~\ref{App:inv_states}. We obtain 
\begin{align}\label{Pexp93543}
    Q = 2 (1-\eta) \lambda_0 (1-\lambda_0)   + \eta (\lambda_0 + \lambda_1 - 2 \lambda_0 \lambda_1) \, .
\end{align}
This sets the range of feasibility for the gain,
\begin{align}
    Q  \in [ Q_{\min}, Q_{\max} ]  \, ,
\end{align}
with
\begin{align}
    Q_{\min} & =  \min{ \{ 2 \lambda_0 (1-\lambda_0)  , \lambda_0 + \lambda_1 - 2 \lambda_0 \lambda_1 \} } \, , \\
    Q_{\max} & =  \max{ \{ 2 \lambda_0 (1-\lambda_0)  , \lambda_0 + \lambda_1 - 2 \lambda_0 \lambda_1 \} } \, .
\end{align}
In an experimental implementation, $Q$ can be estimated from the data, from which one in turn determines $\eta$,
\begin{align}\label{qfix}
    \eta = \frac{Q - 2 \lambda_0(1-\lambda_0)}{(1-2\lambda_0)(\lambda_1-\lambda_0)}
\end{align}

\begin{figure}[t!]
\centering
\includegraphics[width=0.95\columnwidth]{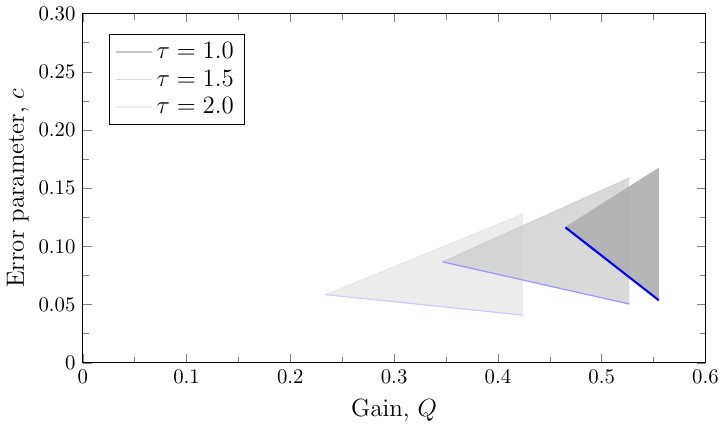} 
\caption{\textbf{Feasible key rate region}: The shaded regions illustrates feasible values for $Q$ and $c$ that are compatible with our model with up to one photon on Bob side for different detector threshold values $\tau$. The feasible region for $\tau=1$ is illustrated in darkest gray. Lighter shades of gray are for $\tau=1.5$ and $\tau=2.0$. The blue lines on the bottom boundaries correspond to passive attacks (for varying values of the loss factor $\eta$), in which case the key rate reduces to Eq.~\eqref{eqn:asymrate_passive}.}
\label{fig:reg}
\end{figure}

Similarly, the parameters $c$ reads
\begin{align}\label{cexp0987}
    c & = (1-\eta) \frac{1}{2} \lambda_0 (1-\lambda_0) \nonumber \\
    & \phantom{=}~ + \eta \left( \frac{2f+1}{6} (1-\lambda_1) \lambda_0 
        + \frac{1-f}{3} (1-\lambda_0) \lambda_1 \right) \, .
\end{align}
Given $\eta$, the range of $c$ is given by
\begin{align}
    c \in \left[ c_{\min} , c_{\max} \right]
\, ,
\end{align}
with 
\begin{align}
c_{\min} & = (1-\eta) \frac{\lambda_0 (1-\lambda_0)}{2}  
    + \eta \frac{ \lambda_0 (1-\lambda_1) }{2} \, , \\
c_{\max} & = 
        (1-\eta) \frac{\lambda_0 (1-\lambda_0)}{2}  
    + \eta \left( \frac{ \lambda_0 (1-\lambda_1) }{6}  
        + \frac{ \lambda_1 (1-\lambda_0) }{3}  \right) \, .
\end{align}
The feasible region for key rates compatible with our protocol is illustrated (for different values of the threshold $\tau$) by the shaded regions in the $Q$-$c$ plane in Fig.~\ref{fig:reg}.

Note that the error parameter $c$ can be estimated from the experimental data, which in turn determines the parameter $f_1$ uniquely,
\begin{align}
 f_1 & = \frac{ 3 (1-\eta) \lambda_0 (1-\lambda_0)}{2\eta (\lambda_1 - \lambda_0) } + \frac{ \lambda_0 +  2\lambda_1 - 3\lambda_0\lambda_1 }{2 (\lambda_1 - \lambda_0)} \nonumber \\
& \phantom{=}~ - \frac{ 3 c}{\eta (\lambda_1 - \lambda_0)}
\, .
\end{align}

In conclusion, the experimental estimates $c$ and $Q$ completely determine the state with no minimisation required to compute the key rate. It remains to compute the relative entropy and hence the rate for given values of these two parameters. The asymptotic rate can then be written as
\begin{align}
    r = (1-\eta) D[\rho_{1:0}^\text{(inv)}] + \eta D[\rho_{1:1}^\text{(inv)}(f_1)] - Q \, h_2(E) \, ,
\label{eq:asymrate_1photon}    
\end{align}
with $E = 2c/Q$. By using the above expressions for $\eta$ and $f_1$, the key rate is entirely determined by the experimental estimates of $Q$ and $c$. In Fig.~\ref{fig:reg}, the blue line at the bottom boundary corresponds to the pure-loss channel, in which case the key rate reduces to Eq.~\eqref{eqn:asymrate_passive}.

\begin{figure}[t!]
\centering
\includegraphics[width=0.95\columnwidth]{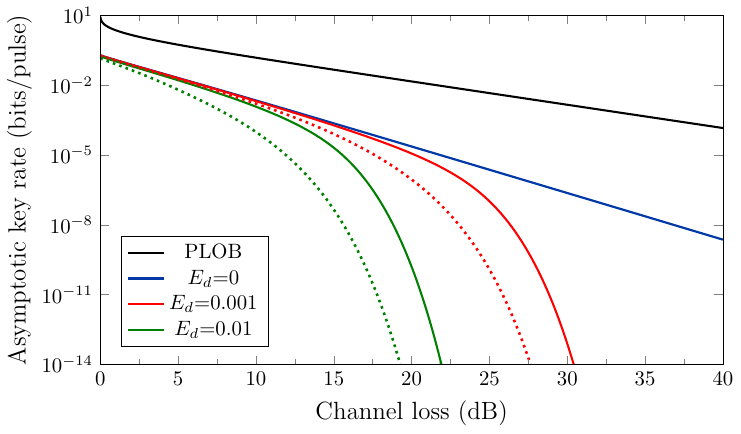} 
\caption{\textbf{Protocol comparison}:
Comparison of asymptotic rates (bits/pulse) vs loss (dB) according to our theory (solid lines) and to the theory of Qi \cite{Qi2021_PRA} (dotted lines), for different values of the error probability, $E_d$. The black line on the top of the figure corresponds to the PLOB repeaterless bound~\cite{PLOB}.}
\label{fig:comp}
\end{figure}

Our results can be directly compared with those of Qi in Ref.~\cite{Qi2021_PRA}. We first need to recall that Qi introduced a model of virtual detectors to provide an upper bound on the key rate as a function of the detector misalignment, quantified by the parameter $E_d$. Leveraging this virtual detection model, the QBER corresponding to Bob detecting a single-photon is~\cite{Qi2021_PRA}
\begin{align}
E_1^\text{Qi} = \frac{(E_d \tau + 1)e^{-\tau} - (\tau+1)e^{-2\tau}}{(\tau+2)e^{-\tau} - 2 (\tau+1)e^{-2\tau}}.
\end{align}
The key rate in Ref.~\cite{Qi2021_PRA} can then be written using our notation, 
\begin{align}
    r_\text{Qi} =
    Q_0 + Q_1 \left( 1-h_2(E_d)\right)  
    - Q \, h_2(E) \, ,
    \label{eq:Qi_result_rate}
\end{align}
with
\begin{align}
    Q_0 & = 2 (1-\eta) \lambda_0 (1-\lambda_0)  \, , \\
    Q_1 & = \eta( \lambda_0 + \lambda_1 - 2 \lambda_0 \lambda_1 ) \, , \\
    Q & = Q_0 + Q_1 \, , \\
    E & = \frac{ Q_0 E_0 + Q_1 E_1^\text{Qi}}{ Q } 
    = \frac{ Q_0 /2 + Q_1 E_1^\text{Qi}}{ Q } \, .
\label{eq:Qi_result}    
\end{align}
Now, to compare the key rate in Eq.~\eqref{eq:Qi_result_rate} with our formalism in Eq.~\eqref{eq:asymrate_1photon}, we determine an expression for $f_1$ using our expression for the QBER conditioned on Bob receiving a single photon 
\begin{align}
E_1 = \frac{2c_1}{Y_1} = \frac{1}{3} \frac{  \lambda_0 + 2 \lambda_1 - 3 \lambda_0 \lambda_1 - 2 f_1 (\lambda_1-\lambda_0) }{\lambda_0 + \lambda_1 - 2\lambda_0 \lambda_1} \, .
\end{align}
By equating this to $E_1^\text{Qi}$, we find
\begin{align}
f_1 = 1 - \frac{3 E_d}{2},
\end{align}
which is independent of the detector threshold, $\tau$. Note that for $E_d=0$ we obtain $f_1=1$ and our key rate recovers the rate for passive attacks and matches the result of Qi. For non-zero $E_d$, Fig.~\ref{fig:comp} illustrates a comparison of the rates achieved with our formalism with Ref.~\cite{Qi2021_PRA}. Our theory provides higher rates and can tolerate higher channel losses.

%


\section{Controlled minimisation of the relative entropy}
\label{sec:minimisation_rel_ent}

\noindent
In general, there is no guarantee that the state obtained by Bob involves only one photon or even a bounded number of photons. However, Bob can estimate the photon number distribution using the output of heterodyne detection~\cite{Qi:het}. In practice, only a few parameters $P_j$ in the expansion (\ref{symstate0}) will be realistically estimated with a reasonable small error, say from $j=0$ up to $j=k$. The limited information on the parameters $P_j$ is still useful to obtain a lower bound on the relative entropy. In fact from Eq.~(\ref{Dinv}) we obtain
\begin{align}
D[\rho_{AB}^\text{(inv)}] \geq P_0 D[\rho_{1:0}^\text{(inv)} ] + \sum_{j=1}^k P_j D[\rho_{1:j}^\text{(inv)} (f_j) ] \, .
\end{align}

Since Alice and Bob can estimate the parameters $Q$, $c$, and $P_j$ for $j$ between $0$ and $k$ from their experimental data, a lower bound on the relative entropy is obtained by solving the constrained minimisation:
\begin{align}\label{eqn:eqiv_min}
D[ \rho_{AB}^\text{(inv)}] \geq  P_0 D[ \rho_{1:0}^\text{(inv)} ] + \min_{f_1,\dots,f_k}  \sum_{j=1}^k  P_j  D[ \rho_{1:j}^\text{(inv)} (f_j) ] \, ,
\end{align}
where the minimisation is subject to the constraint
\begin{align}\label{eqn:eqiv_min_con}
P_0 c_{0} + \sum_{j=1}^k  P_j  c_{j}(f_j) \leq c  \, .
\end{align}
Since we expect the relative entropy to decrease monotonically with increasing $c$, we may replace this inequality with an equality. Note that the optimisation in Eq.~\eqref{eqn:eqiv_min} is over $k$ parameters $f_j \in [0,1]$, for $j=1,\dots,k$. Therefore, the complexity of the optimisation is reduced from quadratic to linear in the photon number cutoff.

Alternatively, one can use the estimated QBER in the constrained minimisation instead of the parameter $c$, yielding
\begin{align}\label{eqn:eqiv_min_2}
2 \, \frac{ P_0
c_{0} 
+ \sum_{j=1}^k 
P_j
c_{j}(f_j) }{ 
Q_{(k)} 
} \leq E
\, ,
\end{align}
where $Q_{(k)}$ is an upper bound for $Q$. As shown in Appendix~\ref{App:ranges}, a suitable upper bound is
\begin{align}
Q_{(k)} = 
\sum_{j=0}^k 
P_j Y_{j} 
+ 
\left( 1 - \sum_{j=0}^k P_j \right)
Y_{k+1}  \, .
\end{align}


\section{Application: assessing the robustness to electronic noise}\label{sec:Gauss}

\noindent
We apply our theory to assess the robustness of the hybrid protocol against electronic noise in heterodyne detection. Electronic noise is one of the most significant challenges for QKD protocols based on coherent detection. We model the electronic noise as Gaussian noise with zero mean and variance $N$, with the following representation as a quantum channel acting on each mode of the field:
\begin{align}\label{noisech}
\rho \to \int \frac{d^2\alpha}{\pi N} e^{-|\alpha|^2/N} \mathcal{D}(\alpha) \rho \mathcal{D}(\alpha)^\dag  \, ,
\end{align} 
where $\mathcal{D}(\alpha)$ is the displacement operator. Note that when applied to the two modes received by Bob, this map preserves the $(U \otimes U^*)$ symmetry.

Overall, we model the communication channel from Alice to Bob as a Gaussian channel obtained by first applying a pure-loss channel of transmissivity $\eta$, followed by mode-wise application of the channel in Eq.~(\ref{noisech}). In the Heisenberg picture, this is described by the map
\begin{align}
b_H & \to \sqrt{\eta} \, b_H + \sqrt{1-\eta} \, e_H + z \, , \\
b_V & \to \sqrt{\eta}\,  b_V + \sqrt{1-\eta} \, e_V + z^* \, ,
\end{align}
where $z$ is a circularly symmetric, complex-valued Gaussian random variable with zero mean and variance $N$.

Using the expansion of the displacement operator in the number bases~\cite{Cahill1969_PR} (for $m \geq n$)
\begin{align} \label{Dnumber}
    \langle m | \mathcal{D}(\alpha) | n \rangle = \sqrt{ \frac{n!}{m!} } \, \alpha^{m-n} e^{-|\alpha|^2/2} L_n^{(m-n)}(|\alpha|^2) \, ,
\end{align}
where $L_n^{(m-n)}$ denotes the Laguerre polynomials, we are able to compute the invariant states. We truncate the Hilbert space to three photons on Bob side. By repeated applications of Eq.~(\ref{Dnumber}) we obtain (details in Appendix~\ref{App:Gauss})
\begin{align}
    \rho_{AB}^\text{(inv)} = P_{0} \rho_{1:0}^\text{(inv)} + \sum_{j=1}^{3} P_{j} \rho_{1:j}^\text{(inv)} (f_j) \, ,
\end{align}
with
\begin{align}
f_1 &= \frac{2 \eta +N^2-\eta  N + N}{2 (\eta +2 N^2-2 \eta  N + 2 N)}\, , \\
f_2 &= \frac{3 \eta +N^2-\eta  N + N}{3 (\eta + N^2 - \eta  N + N)}\, , \\
f_3 &= \frac{3 \left(4 \eta + N^2 - \eta  N + N\right)}{4 (3 \eta + 2 N^2 - 2 \eta  N + 2 N)} \, ,
\end{align}
and $P_0$, $P_1$, $P_2$, $P_3$ are given by the formula
\begin{align}
P_j & = \frac{\eta}{N+1} \sum_{m=0}^j p_m \left( \frac{N}{N+1} \right)^{j-m} \nonumber \\
& \phantom{=}~ + (j+1) \frac{1-\eta}{(N+1)^2} \left( \frac{N}{N+1} \right)^j
\end{align}
where
\begin{align}
    p_m := \left\{
\begin{array}{lcr}
\frac{N}{(N+1)^2} & \,\, \text{if} \,\, & m=0 \\
\frac{1}{N+1} \left( \frac{N}{N+1} \right)^{m} \frac{m + N^2}{N(N+1)} & \,\, \text{if} \,\, & m \geq 1 
\end{array}
    \right.
\end{align}

\begin{figure}[t!]
\centering
\includegraphics[width=0.95\columnwidth]{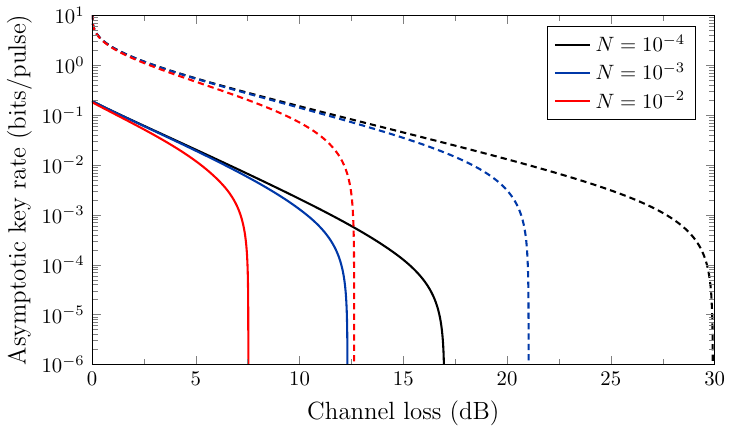} 
\caption{Comparison of asymptotic key rates as a function of loss (dB) for different excess noise, quantified by the noise variance $N$. Solid lines: our lower bound for the hybrid protocol, computed using Eq.~(\ref{3Gauss}) after optimisation of the threshold value $\tau$. Dashed lines: upper bound for continuous-modulation CV QKD, obtained from the reverse coherent information in Eq.~(\ref{cv}).} 
\label{fig:electronic_noise}
\end{figure}%

From this we obtain a lower bound on the relative entropy:
\begin{align}
D[ \rho_{AB}^\text{(inv)} ]  \geq P_0 D[ \rho_{1:0}^\text{(inv)} ] + \sum_{j=1}^3 P_j D[\rho_{1:j}^\text{(inv)} (f_j) ] \, .
\end{align}
Note that the function $Q h_2[2c/Q]$ is monotonically increasing with both $Q$ and $c$. Therefore, an upper bound on the error correction leak is obtained from upper bounds on $Q$ and $c$ (these upper bounds are needed only for our numerical simulation; in any experimental implementation the values of $Q$ and $c$ can be directly estimated from the data).

As discussed in Appendix~\ref{App:ranges}, suitable upper bounds are
\begin{align}
Q_{(3)} & = 
\sum_{j=0}^3 P_j 
Y_{j} 
+ 
\left(
1 - \sum_{j=0}^3 P_j
\right)
Y_{4} \, , \\
c_{(3)} & = 
P_0 c_{0} 
+ \sum_{j=1}^3 P_j 
c_{j}(f_j) 
+ \left(
1 - \sum_{j=0}^3 P_j
\right) \frac{1-\lambda_0}{2}
\, .
\end{align}

In conclusion, we obtain the following lower bound on the asymptotic key rate:
\begin{align}\label{3Gauss}
r \geq P_0 D[\rho_{1:0}^\text{(inv)} ] + \sum_{j=1}^3 P_j D[\rho_{1:j}^\text{(inv)} (f_j) ] - Q_{(3)}  h_2\left[\frac{2c_{(3)}}{Q_{(3)}}\right]  \, ,
\end{align}
this rate is expected to be tight if the variance $N$ of the Gaussian noise is not too large, which in turn implies a small value for  the probability $(1 - \sum_{j=0}^3 P_j)$.

The key rate is illustrated in Fig.~\ref{fig:electronic_noise}. The hybrid protocol is sensitive to excess noise in the detector with $N=10^{-6}$ closely approximating the ideal scenario of no electronic noise. Suppression of excess noise down to the $10^{-4}$ regime in CV-QKD is possible through carrier frequency switching~\cite{Dong2023_E}. In Fig.~\ref{fig:electronic_noise} we compare the performance of our hybrid protocol with CV QKD. Following Ref.~\cite{RcohI}, an upper bound on the key rate achievable in CV QKD with heterodyne detection and reverse reconciliation is given by the \textit{reverse coherent information},
\begin{align}\label{cv}
    r_{\text{CV}}  \leq \log{\left(\frac{1}{1-\eta}\right)} - g(N) \, , 
\end{align}
where $g(N) := (N+1) \log{(N+1)} - N \log{N}$. Note that for an excess noise of $N=10^{-4}$, our scheme can tolerate losses up to $\sim$17~dB, corresponding to an optical fibre transmission of 85~km. The protocol can therefore deliver high-rate QKD in terrestrial or free-space quantum networks over metropolitan scales.


\section{Conclusions}
\label{sec:end}

\noindent
Security proofs in quantum cryptography are often limited to a specific protocol and generally require a theoretical model that precisely matches the physical devices used in their implementation. Closing the disparity between theory and implementation of DV and CV QKD has therefore been the subject of significant effort~\cite{Ghorai_sym,Matsuura2021_NC,Primaatmaja2022_Q}. Recent numerical approaches have provided easier implementation by enabling reliable calculation of key rates that are robust to both device imperfections and changes in protocol structure~\cite{Coles2012, Coles2016_NC, Winick2018_Q}. An alternative research direction that offers a promising route towards implementation is the development of hybrid QKD protocols that strive to assimilate the best features of both DV and CV protocols \cite{Qi2021_PRA,Primaatmaja2022_Q}. Most notable of these features is better range performance and mature security proofs inherited from DV protocols and the scalability and compatibility with existing telecommunication infrastructures inherited from CV protocols due to the use of coherent detection. 

We explore the security of hybrid BB84 with heterodyne detection proposed by Qi in Ref.~\cite{Qi2021_PRA}, where information is encoded is in discrete variables (e.g.~polarisation), and decoding is by heterodyne detection. This variant offers two additional advantages. First, in contrast to DV QKD, it does not require sifting, as a single decoding measurement applies to both encoding bases. Second, in contrast to CV QKD, it does not require a shared local oscillator. However, this proposal requires a shared reference frame (though a reference-frame free version could be envisaged along the lines of Ref.~\cite{Laing}). 

Compared to the protocol of Qi, we add a symmetrisation step to make the protocol invariant under local LOP transformations of the form $U \otimes U^*$, such that Alice and Bob can randomly change the reference frame at each photon transmission. By exploiting symmetry, our modified protocol takes advantage of invariant states that are block-diagonal in the number basis with reduced complexity. Our modified protocol therefore offers several advantages over previous protocols. First, it enables a simplified security analysis. Second, our use of symmetry allows for semi-analytical expressions for the asymptotic key rate under collective attacks. Finally, it enables an efficient numerical procedure to optimise the secret key rate with quadratic speedup. In particular, we are able to perform an exact numerical optimisation with full control of the error due to finite-dimensional cut-off of the otherwise infinite-dimensional Hilbert space typical of CV QKD protocols.

We apply our theory to a few examples of quantum channels connecting Alice to Bob, including linear loss, passive attacks, and Gaussian noise. Our analysis sheds light on the salient features of hybrid QKD: (1) the study of linear loss shows that the key rate scale as $O(\eta^2)$, where $\eta$ is the attenuation factor, instead of the linear scaling that characterises most DV and CV protocols; (2) when Gaussian noise is introduced the key rate decreases rapidly, even when compared with CV protocols, this suggests that the hybrid approach is not necessarily advantageous in terms of robustness to noise.

Returning to the original motivation of improving the implementation of QKD protocols, our work achieves this by introducing a symmetrised hybrid QKD protocol. Our results pave the way for a number of interesting research questions that may further improve the performance of hybrid protocols. First, our theory can be directly extended to include decoy states. Second, it may be possible to prove that invariant states are optimal without introducing active symmetrisation. Third, it would be interesting to introduced post-selection in the protocol, which may increase the achievable distance, and to explore the \textit{differential detection mode} of Ref.~\cite{Qi2021_PRA}. Finally, our approach may be extended to reference-frame-independent QKD~\cite{Laing}, hence removing the need of maintaining a shared reference frame and paving the way to satellite-based applications. In a broader context, our framework to establish invariant states provides a general utility that can be applied to other use cases, such as semi-device-independent communication protocols.


\section*{Acknowledgments}

\noindent
This work has received funding from the UKNQTP and the Quantum Technology Hub in Quantum Communications (EPSRC Grant EP/T001011/1), the European Union's Horizon Europe research and innovation programme under the project ``Quantum Secure Networks Partnership" (QSNP, grant agreement No~101114043), the European Union's Next Generation EU: PNRR MUR project PE0000023-NQSTI, and the INFN through the project ``QUANTUM''. We warmly thank Marco Lucamarini for inspiring discussions.


\bibliographystyle{apsrev4-2}

\clearpage

\begin{widetext}

\appendix


\section{$(U \otimes U^\ast)$-invariant states} 
\label{App:groupt}

\noindent
Suppose that both Alice's and Bob's Hilbert spaces are endowed with $\SU$ representations. For $U\in\SU$, Alice's space transforms under $U$, and Bob's space under $U^\ast$. \emph{Are there any states of the composite system that are left invariant by these transformations?} If yes, \emph{what is their most general form?} We will informally refer to such states as $(U\otimes U^\ast)$-invariant. The purpose of this appendix is to answer some variations on the theme of the previous questions.


\subsection{Preliminaries}
\label{sec-Prelim}
\noindent
In this section we will recall some basic notions, allowing us to give meaning to the problem in different contexts, and study it in the framework of the representation theory of $\SU$.

\medskip
\label{sec-SumAng}
\begin{paragrapho}[Addition of two spin angular momenta]
Let $\bm{J}_\alpha$ ($\alpha=A,B$) be a spin-$j_\alpha$ angular momentum, with basis $\ket{j_\alpha,m_\alpha}_\alpha$, generating a unitary irreducible
representation $\mathscr D_\alpha$ of $\SU$. The addition of the two angular momenta, denoted as $\bm{J}_A+\bm{J}_B$, is an angular momentum, generating the $j_A \times j_B$ representation $\mathscr D_A\otimes \mathscr D_B\colon U\mapsto \mathscr D_A(U)\otimes \mathscr D_B(U)$, which is unitary and (completely) reducible.

The irreducible invariant subspaces --- i.e., invariant subspaces with no invariant proper subspace; any invariant subspace is a direct sum of irreducible ones --- of $\bm{J}_A+\bm{J}_B$, or, which is the same, of $\mathscr D_A\otimes \mathscr D_B$, are precisely the eigenspaces of $(\bm{J}_A+\bm{J}_B)^2$. The whole space is decomposed into $\min(2j_A,2j_B)=j_A+j_B-\abs{j_A-j_B}$ eigenspaces of $(\bm{J}_A+\bm{J}_B)^2$, labelled by the quantum number $j$, which varies by integer steps from $j_A+j_B$ to $\abs{j_A-j_B}$. On each eigenspace, $\bm{J}_A+\bm{J}_B$ is a spin-$j$ angular momentum, and there is an orthonormal basis $\ket{j,m}$, with  $m$ (the eigenvalue of $J_{A,z}+J_{B,z}$) varying by integer steps from $j$ to $-j$. The vectors $\ket{j,m}$ with all possible values of $j$ and $m$, form an orthonormal basis of the whole space.

A particular basis of this kind can be singled out by the following conditions~\cite{biede}:
\begin{align}\label{eq-CS}
& \ket{j,m} = \sqrt{\frac{(j+m)!}{(j-m)! (2j)!}}\,J_-^{j-m}\ket{j,j} \, ,  \\
\label{eq-PhaseConv}
& {}_A \bra{j_A,m_A} {}_B \langle j_B,m_B | j,m\rangle >0 \,.
\end{align}
Condition~\eqref{eq-CS}, often called \emph{Condon--Shortley phase convention}, yields the standard action of the ladder operators, where the relative phase between $\ket{j,m\pm 1}$ and $J_\pm\ket{j,m}$ is 1. If $\ket{j,m}$ is a basis for a spin-$j$ angular momentum, and Eq.~\eqref{eq-CS} holds, the matrix elements of the associated representation are the corresponding coefficients of the Wigner matrix $D^{(j)}$ --- hereafter, whenever we choose an arbitrary basis for a spin angular momentum, the phase convention~\eqref{eq-CS} will always be assumed. In the present context, condition~\eqref{eq-CS} allows us to obtain each $j$-multiplet $\ket{j,m}$ by repeated applications of the destruction operator on $\ket{j,j}$, hence determining a basis up to an overall phase for each multiplet, which is singled out by condition~\eqref{eq-PhaseConv}. The two conditions together ensure the reality of the Fourier coefficients of the basis vectors, with respect to the product basis. In other words, the transition matrices between the two bases are not only unitary but also orthogonal. The basis vectors determined by the above prescriptions are denoted as $\ket{j_A,j_B;j,m}$, and their Fourier coefficients are called Clebsch--Gordan (CG) coefficients, and denoted as $C(j_A,m_A;j_B,m_B;j,m)$, 
\begin{equation}\label{eq-CG}
\ket{j_A,j_B;j,m}
=\,\sum_{m_A,m_B}\,
C(j_A,m_A;j_B,m_B;j,m)
\ket{j_A,m_A}_A\ket{j_B,m_B}_B.
\end{equation}
It is easy to see that the nontrivial CG coefficients must satisfy the selection rule $m_A+m_B=m$, hence the sum at right-hand side of Eq.~\eqref{eq-CG} actually is over one free index.

The irreducible invariant subspaces of $\mathscr D_A\otimes \mathscr D_B$ --- the $(U\otimes U)$-invariant subspaces --- can be expressed in terms of the basis vectors~\eqref{eq-CG} as
\begin{align}\label{eq-UxU}
V^{(2j+1)}_{2j_A:2j_B} = \mathrm{Span}\left\{\,
\ket{j_A,j_B;j,m}  \,\middle|\;   m=-j,\dots,j\,\right\}\,,
\end{align}
for integer $j=j_A+j_B,\dots,\abs{j_A-j_B}$. The irreducible components of $\mathscr D_A\otimes \mathscr D_B$ are obtained by restricting its action to each irreducible invariant subspace,
\begin{equation}
\mathscr D_A(U)\otimes \mathscr D_B(U) \ket{j_A,j_B;j,m}
= \sum_{m'} D^{(j)}_{m'm} (U) \ket{j_A,j_B;j,m'}.
\end{equation}
\end{paragrapho}

\medskip

\begin{paragrapho}[Complex conjugation in $\SU$]
Complex conjugation is an automorphism (i.e., an isomorphism of the group to itself) of $\SU$. Actually, this is true for any special unitary group, since identities $(AB)^\ast=A^\ast B^\ast$, $(A^\ast)^\ast=A$, $(A^\ast)^\dag=(A^\dag)^\ast$, $\det A^\ast=(\det A)^\ast$, hold for square matrices $A,B$, of arbitrary order.
Crucially, complex conjugation is an \emph{inner} automorphism of $\SU$~\cite{artin},
\begin{equation}\label{eq-inner}
U^\ast =(-\iu\sigma_y) \,U\, (-\iu\sigma_y)^{-1} \,,
\end{equation}
that is, taking the complex conjugate of a $\SU$ matrix is the same as taking its adjoint with respect to $-\iu\sigma_y$, an element of the group. Observe that the $\SO$ representation of $-\iu\sigma_y$ is a rotation of 180 degrees around the $y$ axis, i.e., to an inversion of the $xz$ plane. 

These considerations extends to representations. If $\mathscr D$  is a $\SU$ representation, 
\begin{equation}\label{eq-Dtilde}
\tilde{\mathscr D}\colon U\mapsto \mathscr D(U^\ast)
\end{equation}
is in turn a $\SU$ representation, which is unitary if $\mathscr D$ is. These statements hold for any special unitary group. However, if $\mathscr D$ is a representation of $\SU$, $\mathscr D$ and $\tilde{\mathscr D}$ are isomorphic by Eq.~\eqref{eq-inner},
\begin{equation}\label{eq-Dconj}
\tilde{\mathscr D}(U) = \mathscr D(-\iu\sigma_y)\, \mathscr D(U)\, (\mathscr D(-\iu\sigma_y))^{-1} \,,
\end{equation}
and, in particular, unitarily equivalent if $\mathscr D$ is unitary.

If $\mathscr D$ is generated by a spin-$j$ angular momentum $\bm{J}$, by Eq.~\eqref{eq-Dconj}, $\mathscr{\tilde D}$ is generated by
\begin{equation}
\bm{{\tilde{J}}} = \mathscr D(-\iu \sigma_y)\, \boldsymbol{J} \,(\mathscr D(-\iu\sigma_y))^\dag=\exp(\iu\pi J_y)\, \boldsymbol{J} \,\exp(-\iu\pi J_y)=(-J_x,J_y,-J_z)\,,
\end{equation}
a spin-$j$ angular momentum, related to $\bm{J}$ by a rotation of 180 degrees around the $y$ axis. Moreover, if $\ket{j,m}$ is a basis for $\bm{J}$, $\mathscr D(-\iu\sigma_y)\ket{j, m}$ is a basis for $\bm{{\tilde{J}}}$. 
\end{paragrapho}

\medskip
\label{sec-Schwing}
\begin{paragrapho}[Schwinger angular momentum]
Let $\mathscr H$ be a Hilbert space of two independent bosonic modes, that is, with creation and destruction operators $a^\dag_k$ and $a_k$, such that ($k,\ell=1,2$)
\begin{align}
\left[ a_k , a_\ell \right]=0\,,&&
\left[ a_k^\dag , a_\ell^\dag \right] =0\,,&&
\left[ a_k , a_\ell^\dag \right] = \delta_{k \ell} \, \id\,,
\end{align}
with number operators $N_k= a^\dag_k a_k$, and total number operator $N=N_1+N_2$. The vectors
\begin{equation}\label{eq-nn}
\ket{(n_1,n_2)} =  \frac{{a_1^\dag}^{n_1}{a_2^\dag}^{n_2}}{\sqrt{n_1!n_2!}}\ket{0}
\end{equation}
form an orthonormal basis of joint eigenstates of the number operators, where $\ket{0}=\ket{(0,0)}$ is the vacuum state, $a_k\ket{0}=0$.

The Jordan map~\cite{biede},
\begin{equation}\label{eq-Jordan}
(M_{k\ell})_{k,\ell=1}^2
\mapsto 
\sum_{k ,\ell=1}^2 M_{k \ell} \,a^\dag_k a_\ell
\,,
\end{equation}
is a Lie-algebra homomorphism of the matrix algebra of order 2 to operators on $\mathscr H$, that are bilinear in the creation and destruction operators. Since the Hermitian conjugate of a matrix is mapped to the adjoint of the corresponding operator, Hermitian matrices are mapped to observables. In particular, the spin-\textonehalf\ angular momentum $\bm{\sigma}/2$, with ladder operators $\sigma_\pm/2 = (\sigma_x\pm\iu\sigma_y)/2$, is mapped to the \emph{Schwinger angular momentum}~\cite{biede,saku} $\bm{J}=(J_x,J_y,J_z)$, with ladder operators $J_\pm= J_x\pm\iu J_y$,
\begin{align}\label{eq-Schwinger}
J_+=a_1^\dag a_2
\,,&&
J_-= a_2^\dag a_1
\,,&&
J_z
=\frac{a_1^\dag a_1-a_2^\dag a_2}{2}=\frac{N_1-N_2}{2}\,.
\end{align}

Remarkably, $\bm{J}$ yields all the irreducible representations of the Lie algebra $\su$, and generates a representation $\mathscr D$ yielding all the irreducible representations of the Lie group $\SU$~\cite{biede}.  Indeed, the square of the angular momentum is related to the total number operator (corresponding to the image, under the Jordan map, of the identity matrix) by
\begin{equation}\label{eq-Jsquared}
\boldsymbol{J}^2
=\frac{N}{2} \left(\frac{N}{2}+\id \right)\,.
\end{equation}
As a consequence, the eigenspace of $\bm{J}^2$ relative to the quantum number $j$ coincides with the eigenspace $\mathscr H_{2j}$ of the total number operator $N$ relative to the eigenvalue $2j$, which is spanned by $\ket{(n_1,n_2)}$, with $n_1+n_2=2j$. Moreover, by Eqs.~\eqref{eq-Schwinger}--\eqref{eq-Jsquared}, joint eigenvectors of $N_1$ and $N_2$, and joint eigenvectors of $\bm{J}^2$ and $J_z$ coincide, and the corresponding quantum numbers are related by
\begin{align}\label{eq-quantum}
j=\frac{n_1+n_2}{2}\,,&&m=\frac{n_1-n_2}{2}\,.
\end{align}
Therefore, $\mathscr H_{2j}$ is invariant under $\bm{J}$, and the restriction of $\bm{J}$ to $\mathscr H_{2j}$ is a spin-$j$ angular momentum, with basis
\begin{equation}\label{eq-jmbasis}
\ket{j,m}
=\frac{{a_1^\dag}^{j+m}{a_2^\dag}^{j-m}}{\sqrt{(j+m)!(j-m)!}}\ket{0},
\end{equation}
with $m=-j,\dots, j$, automatically satisfying the phase convention~\eqref{eq-CS}. Then $\SU$ can act on $\mathscr H$ under the unitary representation $\mathscr D$ generated by $\bm{J}$. In particular, its action on $\mathscr H_{2j}$ is
\begin{equation}\label{eq-irrep}
\mathscr D(U)\ket{j,m}=\sum_{m'} D_{m'm}^{(j)}(U)\ket{j,m'}\,.
\end{equation}
In other words, the subrepresentation $\mathscr D^{(j)}\colon U\mapsto \mathscr D(U)\vert_{\mathscr H_{2j}}$, restricting the action of $\mathscr D$ to $\mathscr H_{2j}$, is a spin-$j$ representation on $\mathscr H_{2j}$, generated by the restriction
of $\bm{J}$ to $\mathscr H_{2j}$.

As to complex conjugation, the representation $\mathscr {\tilde D}$ defined by Eq.~\eqref{eq-Dtilde} is generated by the rotated angular momentum $\bm{{\tilde{J}}} = (-J_x,J_y,-J_z)$, the Schwinger angular  momentum associated to the rotated spin-\textonehalf\ angular momentum 
$\bm{{\tilde{\sigma}}}/2 = (-\sigma_x/2,\sigma_y/2,-\sigma_z/2)$. Then $\mathscr H_{2j}$ is invariant under $\mathscr {\tilde D}$, and the spin-$j$ subrepresentations of $\mathscr  D$ and $\mathscr {\tilde D}$ on $\mathscr H_{2j}$ are related by
\begin{equation}\label{eq-RestrVsConj}
\mathscr {\tilde D}{}^{(j)}(U)=\mathscr D^{(j)}(U^\ast)\,.
\end{equation}
\end{paragrapho}

\subsection{Abstract problem}\label{sec-Spin}
\noindent
Let us consider the simple case in which Alice's and Bob's representations are both irreducible, namely, let $\bm{J}_\alpha$ ($\alpha=A,B$)  be a spin-$j_\alpha$ angular momentum on a Hilbert space $\mathscr H_\alpha$, with basis $\ket{j_\alpha,m_\alpha}_\alpha$, generating a representation $\mathscr D_\alpha$ of $\SU$. 
Let's also keep in mind that the relevant case to our real problem will be $j_A=1/2$. While theoretical considerations will generally be carried out for a generic $j_A$, we will generally look for explicit expressions only for $j_A=1/2$.

\subsubsection{General considerations}
\noindent
We know that $\mathscr D_A\otimes \mathscr D_B$ and $\mathscr D_A\otimes \mathscr {\tilde D}_B$ are unitary representations of $\SU$, acting on $\mathscr H_A\otimes \mathscr H_B$, respectively generated by the angular momenta $\bm{J}_A+\bm{J}_B$ and $\bm{J}_A + \bm{{\tilde{J}}}_B$, and unitarily equivalent by Eq.~\eqref{eq-Dconj},
\begin{equation}
\mathscr D_A(U)\otimes \tilde{\mathscr D}_B(U)
= (\id_A\otimes \mathscr D_B(-\iu\sigma_y) )
( \mathscr D_A(U)\otimes\mathscr D_B(U))
( \id_A\otimes \mathscr D_B(-\iu\sigma_y) )^\dag\,.
\label{eq-DxDbarra}
\end{equation}
\medskip
\begin{paragrapho}[$(U\otimes U^\ast)$-invariant subspaces from $(U\otimes U)$-invariant subspaces]
Since $\mathscr D_A\otimes \mathscr D_B$ and $\mathscr D_A\otimes \mathscr {\tilde D}_B$ are related by a unitary transformation, the $(U\otimes U^\ast)$-invariant subspaces are all and only the images of the $(U\otimes U)$-invariant subspace under the unitary operator $\id_A\otimes \mathscr D_B(-\iu\sigma_y)$. Moreover, we are free to multiply this unitary, on the left, by any unitary of the kind $\mathscr D_A(U_0)\otimes \mathscr D_B(U_0^\ast )$, with $U_0\in\SU$. Since we are essentially interested to $j_A=1/2$, a suitable choice, allowing us to move the identity on the higher spin, is $\mathscr U=\mathscr D_A(-\iu\sigma_y)\otimes \id_B$. Incidentally, $\mathscr U$ is the precise analogue of $\id_A\otimes \mathscr D_B(-\iu\sigma_y)$ for $\mathscr {\tilde D}_A\otimes \mathscr D_B$. This fact should not come as a surprise: $(U\otimes U^\ast)$-invariance and $(U^\ast\otimes U)$-invariance coincide, after all. 

By Eq.~\eqref{eq-UxU}, the whole space is decomposed into the irreducible $(U\otimes U^\ast)$-invariant subspaces
\begin{equation}\label{eq-UxUast}
W^{(2j+1)}_{2j_A:2j_B}
=\mathscr U  V^{(2j+1)}_{2j_A:2j_B}=\mathrm{Span}\left\{ \,\mathscr U \ket{j_A,j_B;j,m}
\,\middle|\; 
m = -j,\dots,j\, \right\}\,,
\end{equation}
with $j=j_A+j_B,\dots,\abs{j_A-j_B}$. 
\end{paragrapho}
\medskip
\begin{paragrapho}[
Invariant states from invariant subspaces]
We say that a density operator $\rho$ of the bipartite system is a $(U\otimes U^\ast)$-invariant state if
\begin{align}\label{eq-UxUastInvStDef}
\rho=\mathscr D_A(U)\otimes \mathscr D_B(U^\ast) \,\rho\,
\left(\mathscr D_A(U)\otimes \mathscr D_B(U^\ast)\right)^\dag\,,
\end{align}
for all $U\in\SU$, that is, if $\rho$ commutes with $\mathscr D_A\otimes \mathscr {\tilde D}_B$, or, which is the same, with each component of $\bm{J}_A + \bm{{\tilde{J}}}_B$. 
Now, if $\rho$ is an invariant state, each invariant subspace of $\mathscr D_A\otimes \mathscr {\tilde D}_B$ must be invariant under $\rho$. Consequently, $\rho$ is decomposed into a direct sum of states, each living in different irreducible invariant subspace of $\mathscr D_A\otimes \mathscr {\tilde D}_B$. But then, each such state is a scalar by Schur's lemma~\cite{hall}. As a result, the following states are $(U\otimes U^\ast)$-invariant:
\begin{align}\label{eq-UxUastInvSt}
\rho^{(2j+1)}_{2j_A:2j_B}=\frac{1}{2j+1}\,P^{(2j+1)}_{2j_A:2j_B}\,,
\end{align}
where $P^{(2j+1)}_{2j_A:2j_B}$ is the orthogonal projection onto $W^{(2j+1)}_{2j_A:2j_B}$, and the most general invariant state is a convex combination of states of the kind~\eqref{eq-UxUastInvSt}.
\end{paragrapho}

\subsubsection{Discussion of the case $\frac{1}{2}\times j_B$}

\noindent
Let us assign the role of ``computational basis'' to the product basis $\ket{j_A,m_A}_A \ket{j_B,m_B}_B$, and look for explicit expressions of the $(U\otimes U^\ast)$-invariant states. Since an invariant state is a linear combination of orthogonal projections over irreducible invariant subspaces, and, by Eq.~\eqref{eq-UxUast}, such subspaces are spanned by orthonormal vectors in the form $\mathscr U \ket{j_A,j_B;j,m}$, all we have to do is decompose these vectors on the product basis. To this end, we need the matrix elements of $\mathscr U $, that is, of $\mathscr D_A(-\iu\sigma_y)$, and the CG coefficients for $j_A\times j_B$. 

Hereafter we will show how to handle this problem for the case of our concern, $j_A=1/2$. First of all, our notation for a spin-\textonehalf\ can be simplified by setting $\ket{\pm}= \ket{\tfrac{1}{2},\pm  \tfrac{1}{2}}$. We know that the whole space is decomposed as
\begin{equation}
	\mathscr H_{A}\otimes \mathscr H_{B}
	=V^{(2j_B+2)}_{1:2j_B}\oplus V^{(2j_B)}_{1:2j_B}
	=W^{(2j_B+2)}_{1:2j_B}\oplus W^{(2j_B)}_{1:2j_B}\,,
\end{equation}
that is, into two irreducible $(U\otimes U)$-invariant subspaces, as well as two irreducible $(U\otimes U^\ast)$-invariant subspaces, of dimension $2j_B+2$ and $2j_B$, related by
\begin{align}\label{eq-VW}
W^{(2j_B+2)}_{1:2j_B}=\mathscr U V^{(2j_B+2)}_{1:2j_B}\,,
&&
W^{(2j_B)}_{1:2j_B}=\mathscr U V^{(2j_B)}_{1:2j_B}\,.
\end{align}
The basis vectors of the irreducible $(U\otimes U)$-invariant subspaces are precisely the two $j$-multiplets for $\tfrac{1}{2}\times j_B$ ($j=j_B \pm 1/2$), and are linear combination of at most two product basis vectors,
\begin{multline}
\ket{\tfrac{1}{2},j_B;j_B\pm\tfrac{1}{2},m}=
C\left(\tfrac{1}{2},\tfrac{1}{2};j_B,m-\tfrac{1}{2};j_B\pm\tfrac{1}{2},m\right)\,
\ket{+}_A\,\ket{j_B,m - \tfrac{1}{2}}_B\\
+C\left(\tfrac{1}{2},-\tfrac{1}{2};j_B,m+\tfrac{1}{2};j_B\pm\tfrac{1}{2},m\right)\,
\ket{-}_A\,\ket{j_B,m + \tfrac{1}{2}}_B.
\end{multline}
Due to the phase convention~\eqref{eq-CS}, $\mathscr D_A$ has the same matrix elements as the defining representation; in simple terms,  $\mathscr D_A(U)$ and $U$ can be identified. In particular, $\mathscr D_A(-\iu\sigma_y)\ket{\pm}_A =\pm\ket{\mp}_A$. As a result, the basis vectors of the irreducible $(U\otimes U^\ast)$-invariant subspaces read
\begin{multline}\label{eq-max}
\mathscr U  \ket{\tfrac{1}{2},j_B;j_B\pm\tfrac{1}{2},m}=
C\left(\tfrac{1}{2},\tfrac{1}{2};j_B,m-\tfrac{1}{2};j_B\pm\tfrac{1}{2},m\right)\,
\ket{-}_A\,\ket{j_B,m - \tfrac{1}{2}}_B\\
-C\left(\tfrac{1}{2},-\tfrac{1}{2};j_B,m+\tfrac{1}{2};j_B\pm\tfrac{1}{2},m\right)\,
\ket{+}_A\,\ket{j_B,m + \tfrac{1}{2}}_B.
\end{multline}
The calculation the CG coefficients for $\tfrac{1}{2}\times j_B$ is also rather straightforward, and will be shown below, in order to make this appendix as self-consistent as possible.
Actually, this task will be carried out by first computing the CG coefficients for $j_A\times j_B$ with the two highest value of $j$, and then specializing the results to $j_A=1/2$. Indeed, there would be no advantage in assuming $j_A=1/2$ right from the start; on the contrary, while our math would not get any simpler, the resulting notation would become quite cumbersome.

\medskip

\begin{paragrapho}[Clebsch--Gordan coefficients for $j_A\times j_B$, $j=j_A+j_B$]
Throughout this paragraph, and the next one, $j_A$ and $j_B$ are fixed; therefore we will write $\ket{m_A}\ket{m_B}$ instead of $\ket{j_A,m_A}_A\ket{j_B,m_B}_B$, and $\ket{j,m}$ instead of $\ket{j_A,j_B;j,m}$. We will also set $\hat \jmath=j_A+j_B$.

Let us start with the $\hat \jmath$-multiplet. The eigenspace of $J_{A,z}+J_{B,z}$ relative to $m=\hat\jmath$ is spanned by $\ket{j_A}\ket{j_B}$. Then, by condition~\eqref{eq-PhaseConv} we must set
\begin{equation}
\ket{\hat\jmath,\hat\jmath}=\ket{j_A}\ket{j_B},
\end{equation}
and, by condition~\eqref{eq-CS}, 
\begin{equation}\label{eq-jmax_m}
\ket{\hat\jmath,m}
=\sqrt{\frac{(\hat\jmath+m)!}{(\hat\jmath-m)! (2\hat\jmath)!}}\,
J_-^{\hat\jmath-m}\ket{j_A}\ket{j_B}.
\end{equation}
We first compute
\begin{align}
J_-^{\hat \jmath-m}\ket{j_A}\ket{j_B}& = (J_{A,-}+J_{B,-})^{\hat \jmath-m}\ket{j_A}\ket{j_B}\nonumber\\
&=\;\;\sum_{\mathclap{\substack{m_A,m_B\\m_A+m_B=m}}}\quad
\frac{(\hat \jmath-m)!}{(j_A-m_A)!(j_B-m_B)!} \,
J_{A,-}^{j_A-m_A}\ket{j_A}
J_{B,-}^{j_B-m_B}\ket{j_B}\nonumber\\
&=
\;\;\sum_{\mathclap{\substack{m_A,m_B\\m_A+m_B=m}}}\quad
\left[ \frac{(\hat \jmath-m)!^2(2j_A)! (2j_B)!}{(j_A-m_A)!(j_A+m_A)!(j_B-m_B)!(j_B+m_B)!} \right]^{\frac{1}{2}}\,
\ket{m_A}
\ket{m_B},
\end{align}
and then plug the result into Eq.~\eqref{eq-jmax_m}, to obtain the whole multiplet,
\begin{align}
\ket{\hat\jmath,m}
&=
\;\;\sum_{\mathclap{\substack{m_A,m_B\\m_A+m_B=m}}}\quad\,
\left[ \frac
{(\hat \jmath+m)!(\hat \jmath-m)!(2j_A)!(2j_B)!}
{(2\hat \jmath)!(j_A-m_A)!(j_A+m_A)!(j_B-m_B)!(j_B+m_B)!} 
\right]^{\frac{1}{2}}\,
\ket{m_A}
\ket{m_B}\nonumber\\
&=
\;\;\sum_{\mathclap{\substack{m_A,m_B\\m_A+m_B=m}}}\quad\,
\left[ \frac{\binom{2j_A}{j_A-m_A}\binom{2j_B}{j_B-m_B}} 
{\binom{2\hat \jmath }{\hat \jmath-m}} \right]^{\frac{1}{2}}\,
\ket{m_A}
\ket{m_B}.
\label{eq-jmax}
\end{align}
Observe that the coefficients with $j=j_A+j_B-1$ are non negative.
\end{paragrapho}
\medskip

\begin{paragrapho}[Clebsch--Gordan coefficients for $j_A\times j_B$, $j=j_A+j_B-1$]
As to the $(\hat \jmath-1)$-multiplet, the eigenspace of $J_{A,z}+J_{B,z}$, relative to $m=\hat\jmath-1$, is spanned by $\ket{j_A}\ket{j_B-1}$ and $\ket{j_A-1}\ket{j_B}$, and, by Eq.~\eqref{eq-jmax},
\begin{equation}\label{eq-jmax_jmax-1}
\ket{\hat\jmath,\hat\jmath-1}
=\sqrt{\frac{j_B}{j_A+j_B}} \,\ket{j_A}\ket{j_B-1}
+\sqrt{\frac{j_A}{j_A+j_B}}\, \ket{j_A-1}\ket{j_B}.
\end{equation}
is one of the eigenstates. Therefore, by condition~\eqref{eq-PhaseConv}, we must set
\begin{equation}\label{eq-jmax-1_jmax-1}
\ket{\hat\jmath-1,\hat\jmath-1}
=\sqrt{\frac{j_A}{j_A+j_B}} \,\ket{j_A}\ket{j_B-1}
-\sqrt{\frac{j_B}{j_A+j_B}}\, \ket{j_A-1}\ket{j_B},
\end{equation}
and, by condition~\eqref{eq-CS}, 
\begin{equation}\label{eq-jmax-1_m}
\ket{\hat\jmath-1,m}=
\sqrt{\frac{(\hat\jmath-1+m)!}{(\hat\jmath-1-m)! (2\hat\jmath-2)! \hat\jmath}}
\,\left[\sqrt{j_A}\,J_-^{\hat\jmath-m-1}\ket{j_A}\ket{j_B-1}
-\sqrt{j_B}\,J_-^{\hat\jmath-m-1}\ket{j_A-1}\ket{j_B}\right].
\end{equation}
The generic state of the multiplet is the sum of two terms. We first compute
\begin{align}
J_-^{\hat \jmath-m-1}\ket{j_A}\ket{j_B-1}
&= \left( J_{A,-}+J_{B,-} \right)^{\hat \jmath-m-1}\ket{j_A}\ket{j_B-1}
\nonumber\\
&=
\;\;\sum_{\mathclap{\substack{m_A,m_B \\m_B\le j_B-1\\m_A+m_B=m}}}\quad
\frac{(\hat \jmath-m-1)!}{(j_A-m_A)!(j_B-m_B-1)!} \,
J_{A,-}^{j_A-m_A}\ket{j_A}
J_{B,-}^{j_B-m_B-1}\ket{j_B-1}
\nonumber\\
&=
\;\;\sum_{\mathclap{\substack{m_A,m_B \\m_B\le j_B-1\\m_A+m_B=m}}}\quad
\left[ \frac{(\hat \jmath-m-1)!^2(2j_A)! (2j_B-1)!(j_B-m_B)}
{(j_A-m_A)!(j_A+m_A)!(j_B-m_B-1)!(j_B+m_B)!} \right]^{\frac{1}{2}}\,
\ket{m_A}
\ket{m_B}
\nonumber\\
&=
\;\;\sum_{\mathclap{\substack{m_A,m_B \\m_A+m_B=m}}}\quad
\left[ \frac{(\hat \jmath-m-1)!^2(2j_A)! (2j_B-1)!(j_B-m_B)^2}
{(j_A-m_A)!(j_A+m_A)!(j_B-m_B)!(j_B+m_B)!} \right]^{\frac{1}{2}}\,
\ket{m_A}
\ket{m_B}.
\label{eq-unconstr}
\end{align}
The term in $J_-^{\hat \jmath-m-1}\ket{j_A-1}\ket{j_B}$ is immediately obtained by substituting $A\leftrightarrow B$ in the coefficients of the above linear combination (hence, no contribution can come from $\ket{j_A}$ either). Plugging these results into Eq.~\eqref{eq-jmax-1_m} yields the whole multiplet,
\begin{align}
\MoveEqLeft{\sqrt{\frac{(\hat\jmath-1+m)! j_A}{(\hat\jmath-1-m)! (2\hat\jmath-2)! \hat\jmath}}
\,J_-^{\hat\jmath-m-1}\ket{j_A}\ket{j_B-1}}
\nonumber\\
&=
\;\;\sum_{\mathclap{\substack{m_A,m_B \\m_A+m_B=m}}}\quad
\left[ \frac{(\hat \jmath-m-1)!(\hat \jmath-m+1)!(2j_A)! (2j_B-1)! j_A(j_B-m_B)^2 }
{(2\hat\jmath-2)! (j_A-m_A)!(j_A+m_A)!(j_B-m_B)!(j_B+m_B)!\hat \jmath} \right]^{\frac{1}{2}}\,
\ket{m_A}
\ket{m_B}
\nonumber\\
&=
\;\;\sum_{\mathclap{\substack{m_A,m_B\\m_A+m_B=m}}}\quad
\left[ \frac{\binom{2j_A}{j_A-m_A}\binom{2j_B}{j_B-m_B}}
{\binom{2\hat \jmath -2}{\hat \jmath-m-1}} \right]^{\frac{1}{2}}
\,\frac{ j_A(j_B-m_B)}{\sqrt{2 j_A j_B \hat \jmath}} \,
\ket{m_A}
\ket{m_B}.
\label{eq-partA}
\end{align}
The term in $J_-^{\hat \jmath-m-1}\ket{j_A-1}\ket{j_B}$ is obtained by substituting $A\leftrightarrow B$ in the coefficients of the above linear combination. Plugging these results into Eq.~\eqref{eq-jmax-1_m} yields the whole multiplet,
\begin{equation}\label{eq-jmax-1}
\ket{\hat\jmath-1,m}=
\;\;\sum_{\mathclap{\substack{m_A,m_B\\m_A+m_B=m}}}\quad
\left[ \frac{\binom{2j_A}{j_A-m_A}\binom{2j_B}{j_B-m_B}}
{\binom{2\hat \jmath -2}{\hat \jmath-m-1} } \right]^{\frac{1}{2}}
\,\frac{j_B m_A-j_Am_B}{\sqrt{2 j_A j_B \hat \jmath}} \,
\ket{m_A}
\ket{m_B}.
\end{equation}
The coefficients with $j=j_A+j_B-1$ have the sign of $m_A/j_A-m_B/j_B$, hence can be negative.

\end{paragrapho}
\medskip

\begin{paragrapho}[Clebsch--Gordan coefficients for $\tfrac{1}{2}\times j_B$]\label{sec-1/2}

By Eq.~\eqref{eq-jmax}, the nontrivial CG coefficients with $j=j_B+1/2$ are in the form
\begin{align}
C\left(\tfrac{1}{2},\pm\tfrac{1}{2}; j_B, m \mp \tfrac{1}{2};j_B+\tfrac{1}{2},m\right)
&=\binom{1}{\frac{1}{2}\mp \frac{1}{2}}^{\frac{1}{2}}
\binom{2j_B}{j_B-m \pm \frac{1}{2}} ^{\frac{1}{2}}
\binom{2 j_B+1 }{j_B-m+\frac{1}{2}}^{-\frac{1}{2}}\nonumber\\
&=\sqrt{\frac{1}{2}\pm\frac{ m}{2j_B+1}}\,,
\end{align}
for $m=j_B+ 1/2,\dots,-j_B - 1/2$. They are all non negative; for $m=\pm(j_B+ 1/2)$ one sign choice yields 0 (and the other 1, as it should be); for any other value of $m$, both choices lead to nontrivial coefficients. 
Likewise, by Eq.~\eqref{eq-jmax-1}, the nontrivial CG coefficients with $j=j_B-1/2$ are in the form
\begin{align}
	C\left(\tfrac{1}{2},\pm\tfrac{1}{2}; j_B, m \mp \tfrac{1}{2};j_B-\tfrac{1}{2},m\right)
	&=\pm
	\binom{1}{\frac{1}{2}\mp \frac{1}{2}}^{\frac{1}{2}}
	\binom{2j_B}{j_B-m \pm \frac{1}{2}} ^{\frac{1}{2}}
	\binom{2 j_B - 1 }{j_B-m-\frac{1}{2}}^{-\frac{1}{2}}
	\frac{j_B\mp m+\tfrac{1}{2}}{\sqrt{2j_B(2j_B+1)}}\nonumber\\
	&=\pm\sqrt{\frac{1}{2}\mp\frac{m}{2j_B+1}}\,,
\end{align}
for $m=j_B -1/2,\dots,-j_B + 1/2$. They are all non vanishing, and have the sign of $m_A=\pm 1/2$.

As a result, the $(j_B+1/2)$-multiplet is made up of two separable states,
\begin{equation}\label{eq-jmaxSepa}
\ket{\tfrac{1}{2},j_B;j_B+\tfrac{1}{2},\pm( j_B+\tfrac{1}{2})}
=\ket{\pm}_A\ket{j_B,\pm j_B}_B,
\end{equation}
corresponding to the highest and lowest values of $J_{A,z}+J_{B,z}$, and the $2 j_B$ entangled states
\begin{equation}\label{eq-jmaxEnt}
\ket{\tfrac{1}{2},j_B;j_B+\tfrac{1}{2},m}=
\sqrt{\frac{1}{2}+\frac{ m}{2j_B+1}}\,\ket{+}_A\ket{j_B,m-\tfrac{1}{2}}_B
+
\sqrt{\frac{1}{2}-\frac{ m}{2j_B+1}}\,\ket{-}_A\ket{j_B,m+\tfrac{1}{2}}_B,
\end{equation}
with $m= j_B- 1/2,\dots,- j_B + 1/2$. These $2j_B+2$ states together generate $V^{(2j_B+2)}_{1:2j_B}$.
On the other hand, the $(j_B-1/2)$-multiplet, is made up of 
\begin{equation}\label{eq-jmax-1_m1/2}
	\ket{\tfrac{1}{2},j_B;j_B-\tfrac{1}{2},m}=
	\sqrt{\frac{1}{2}-\frac{m}{2j_B+1}} \ket{+}_A\ket{j_B,m-\tfrac{1}{2}}_B
	-
	\sqrt{\frac{1}{2}+\frac{m}{2j_B+1}}\ket{-}_A\ket{j_B,m+\tfrac{1}{2}}_B,
\end{equation}
with $m= j_B- 1/2,\dots,- j_B + 1/2$. These $2 j_B$ entangled states generate $V^{(2j_B)}_{1:2j_B}$.

If $\ket{j,\pm(j+1)}=0$ is understood, Eqs.~\eqref{eq-jmaxSepa}--\eqref{eq-jmaxEnt}, and~\eqref{eq-jmax-1_m1/2} can be put together, by saying that $V^{(2j_B+1\pm1)}_{1:2j_B}$ is spanned by the $2j_B+1\pm 1$ orthonormal vectors
\begin{equation}\label{eq-Vsumma}
\ket{\tfrac{1}{2},j_B;j_B\pm\tfrac{1}{2},m}=
\sqrt{\frac{1}{2}\pm\frac{m}{2j_B+1}}\ket{+}_A\ket{j_B,m -\tfrac{1}{2}}_B
\pm\sqrt{\frac{1}{2}\mp\frac{m}{2j_B+1}}\ket{-}_A\ket{j_B,m+\tfrac{1}{2}}_B,
\end{equation}
with $m= j_B \pm 1/2,\dots,- j_B \mp 1/2$.

\end{paragrapho}
\medskip
\begin{paragrapho}[Irreducible $(U\otimes U^\ast)$-invariant subspaces for $\tfrac{1}{2}\times j_B$]
By Eqs.~\eqref{eq-VW} and~\eqref{eq-Vsumma}, $W^{(2j_B+1\pm1)}_{1:2j_B}$ is generated by the $2j_B+1\pm 1$ orthonormal vectors
\begin{equation}\label{eq-Wsumma}
\mathscr U\ket{\tfrac{1}{2},j_B;j_B\pm\tfrac{1}{2},m}=
\sqrt{\frac{1}{2}\pm\frac{m}{2j_B+1}}\ket{-}_A\ket{j_B,m -\tfrac{1}{2}}_B
\mp\sqrt{\frac{1}{2}\mp\frac{m}{2j_B+1}}\ket{+}_A\ket{j_B,m+\tfrac{1}{2}}_B,
\end{equation}
with $m= j_B \pm 1/2,\dots,- j_B \mp 1/2$.
Specifically, $W^{(2j_B+2)}_{1:2j_B}$ is spanned by the $2j_B+2$ vectors
\begin{equation}\label{eq:W}
\begin{gathered}
\ket{-}_A\ket{j_B, j_B}_B,\\
\sqrt{\frac{2j_B}{2j_B+1}}\ket{-}_A\ket{j_B,j_B-1}_B
-\sqrt{\frac{1}{2j_B+1}}\ket{+}_A\ket{j_B,j_B}_B,\\
\quad\vdots\\
\sqrt{\frac{1}{2j_B+1}}\ket{-}_A\ket{j_B,-j_B}_B
-\sqrt{\frac{2j_B}{2j_B+1}}\ket{+}_A\ket{j_B,-j_B+1}_B,\\
-\ket{+}_A\ket{j_B, -j_B}_B,
\end{gathered}
\end{equation}
and $W^{(2j_B)}_{1:2j_B}$ by the $2j_B$ vectors
\begin{equation}\label{eq:w}
\begin{gathered}
\sqrt{\frac{1}{2j_B+1}}\ket{-}_A\ket{j_B,j_B-1}_B
+\sqrt{\frac{2j_B}{2j_B+1}}\ket{+}_A\ket{j_B,j_B}_B,\\
\quad\vdots\\
\sqrt{\frac{2j_B}{2j_B+1}}\ket{-}_A\ket{j_B,-j_B}_B
+\sqrt{\frac{1}{2j_B+1}}\ket{+}_A\ket{j_B,-j_B+1}_B.
\end{gathered}
\end{equation}
The above expressions allow us to write the orthogonal projection over each invariant subspace, hence, by Eq.~\eqref{eq-UxUastInvSt}, the corresponding invariant state.
\end{paragrapho}

\subsection{Actual problem}\label{sec-Inv}

\noindent
Let us now consider two systems, each of two independent photonic modes, namely, the horizontal and the vertical polarization. Alice's system is described by a Hilbert space $\mathscr H_A$, with canonical bosonic annihilation and creation operators $\{ a_k,a^\dag_k\}$ for the optical modes $k=H,V$, determining the number operators $N_A^k$, the total number operator $N_A=N_A^H+N_A^V$, and the standard basis $\ket{(n_A^H,n_A^V)}_A$, as well as the Schwinger angular momentum $\bm{J}_A$, with basis $\ket{j_A,m_A}_A$, and generating the $\SU$ representation $\mathscr D_A$. Likewise for Bob. 

We look for the $(U\otimes U^\ast)$-invariant states --- defined as in Eq.~\eqref{eq-UxUastInvStDef} --- of the composite system, such that Alice's total number is 1, and Bob's is $n$ (an arbitrary, but fixed number). 

\subsubsection{General considerations}
\noindent
This problem is just a special case of the one we studied in Sec.~\ref{sec-Spin}. Indeed, in a Hilbert space of two independent bosonic modes, there exist a correspondence between eigenstates and eigenvalues of number operators, on the one hand, and Schiwnger angular momentum, on the other. In particular, Eqs.~\eqref{eq-nn}, and~\eqref{eq-quantum}--\eqref{eq-jmbasis}, yield $j=n/2$, $m=n^H-n/2$, and $\ket{j,m}=\ket{(n^H,n-n^H)}$ (entailing $\ket{+}=\ket{H}$, and $\ket{-}=\ket{V}$). With this in mind, a bipartite state such that Alice's total number is 1, and Bob's is $n$, lives in the finite-dimensional tensor product space $\mathscr H_{A,1}\otimes \mathscr H_{B,n}$, where Alice's and Bob's representations are both irreducible, with spin $1/2$ and $n/2$, respectively. But then such space is decomposed into two $(U\otimes U^\ast)$-irreducible invariant subspaces,
\begin{equation}
	\mathscr H_{A,1}\otimes \mathscr H_{B,n}
	= W^{(n+2)}_{1:n}\oplus W^{(n)}_{1:n}\,,
\end{equation}
and the most general invariant state is in the form of a convex combination, with one real parameter $ f\in [0,1]$, of the corresponding invariant states,
\begin{equation}\label{eq-invSt}
	\rho_{1:n}^\text{(inv)}(f)
	=(1-f)\rho^{(n+2)}_{1:n} + f\rho^{(n)}_{1:n}
	=\frac{1-f}{n+2}\,P^{(n+2)}_{1:n} +\frac{f}{n}\,P^{(n)}_{1:n} \,,
\end{equation}
where $W^{(2j+1)}_{2j_A:2j_B}$, $\rho^{(2j+1)}_{2j_A:2j_B}$, and $P^{(2j+1)}_{2j_A:2j_B}$ have the same meaning as in Eqs.~\eqref{eq-UxUast} and~\eqref{eq-UxUastInvSt}.  

Let us now turn our attention to the problem of expressing the irreducible invariant subspaces on the product basis,
\begin{equation}\label{eq-tensorbasis}
\ket{H}_A\ket{(n,0)}_B ,\,
\dots,\,
\ket{H}_A\ket{(0,n)}_B,\,
\ket{V}_A\ket{(n,0)}_B ,\,
\dots,\,
\ket{V}_A\ket{(0,n)}_B.
\end{equation}
By Eq.~\eqref{eq:W}, the subspace $W_{1:n}^{(n+2)}$ is spanned by the $n+2$ orthonormal states
\begin{equation}\label{eq-SuExpl}
\begin{gathered}
\ket{V}_A\ket{(n,0)}_B,\\
\sqrt{\frac{n}{n+1}}\ket{V}_A\ket{(n-1,1)}_B
-\sqrt{\frac{1}{n+1}}\ket{H}_A\ket{(n,0)}_B,\\
\vdots\\
\sqrt{\frac{1}{n+1}}\ket{V}_A\ket{(0,n)}_B
-\sqrt{\frac{n}{n+1}}\ket{H}_A\ket{(1,n-1)}_B,\\
-\ket{H}_A\ket{(0,n)}_B.
\end{gathered}
\end{equation}
And by Eq.~\eqref{eq:w} the subspace $W_{1:n}^{(n)}$ is spanned by the $n$ orthonormal states
\begin{equation}\label{eq-GiuExpl}
\begin{gathered}
\sqrt{\frac{1}{n+1}}\ket{V}_A\ket{(n-1,1)}_B
+\sqrt{\frac{n}{n+1}}\ket{H}_A\ket{(n,0)}_B,\\
\vdots\\
\sqrt{\frac{n}{n+1}}\ket{V}_A\ket{(0,n)}_B
+\sqrt{\frac{1}{n+1}}\ket{H}_A\ket{(1,n-1)}_B.
\end{gathered}
\end{equation}

\subsubsection{Explicit solutions for the simplest cases}
\noindent
Hereafter, we will give a glance at the cases in which Bob has $1$, $2$, or $3$ photons. Tensor products will be written as Kronecker products. Matrix representations for linear operators on $\mathscr H_{A,1}\otimes \mathscr H_{B,n}$ shall always be understood with respect to the product basis, ordered as in equation~\eqref{eq-tensorbasis} --- i.e., first by decreasing values of $n_A^H$, then by decreasing values  of $n_B^H$.
We will also emphasize the block structure brought about by the Kronecker product with respect to the ordered basis $(\ket{H}_A , \ket{V}_A )$ of $\mathscr H_{A,1}$. In this way, the density matrix is partitioned into four square blocks, of order $n+1$, and its partial trace on Bob's system is obtained by replacing each block with its own trace.
\medskip
\begin{paragrapho}[Case $n=1$] The $4$-dimensional space $\mathscr H_{A,1}\otimes \mathscr H_{B,1}$, consisting of the states with one photon in each system, is decomposed into invariant subspaces of dimension 3 and 1,
\begin{equation}
\mathscr H_{A,1}\otimes \mathscr H_{B,1}=W_{1:1}^{(3)}\oplus W_{1:1}^{(1)}\,.
\end{equation}
The subspace $W_{1:1}^{(3)}$ is spanned by the orthonormal states
\begin{equation}
\ket{V}_A\ket{H}_B,\;
\frac{1}{\sqrt 2}\bigl(\ket{V}_A\ket{V}_B-\ket{H}_A\ket{H}_B\bigr),\;
-\ket{H}_A\ket{V}_B,
\end{equation}
and corresponds to the invariant state
\begin{align}
\rho_{1:1}^{(3)}
&=\frac{1}{3}\left(\begin{BMAT}(b)[1pt]{cc1cc}{cc1cc}
1/2& 0 & 0 & -1/2\\
0 & 1 & 0 & 0 \\
0 & 0 & 1 & 0 \\
-1/2& 0 & 0 & 1/2
\end{BMAT}\right).
\end{align}
The invariant pure state spanning $W_{1:1}^{(1)}$, $(\ket{HH}+\ket{VV})/{\sqrt 2}$,
yields the density matrix
\begin{equation}
\rho_{1:1}^{(1)}
=\left(\begin{BMAT}(b)[1pt]{cc1cc}{cc1cc}
\mathmakebox[\widthof{$-1/2$}]{
1/2
}& 0 & 0 & 1/2\\
0 & 0 & 0 & 0 \\
0 & 0 & 0 & 0 \\
1/2& 0 & 0 & 1/2
\end{BMAT}\right).
\end{equation}
A generic invariant state is a convex combination of $\rho_{1:1}^{(3)}$ and $\rho_{1:1}^{(1)}$, depending on $f \in [0,1]$,
\begin{equation}\label{inv_11}
\rho_{1:1}^\text{(inv)}(f) 
=(1-f)\,\rho_{1:1}^{(3)} + f \,\rho_{1:1}^{(1)}
 = \frac{1}{6}
\left(\begin{BMAT}(b)[1pt]{cc1cc}{cc1cc}
2f-1& 0 & 0 & 4f-1\\
0 & 2(1-f) & 0 & 0 \\
0 & 0 & 2(1-f) & 0 \\
4f-1 & 0 & 0 &2f-1 
\end{BMAT}
\right).
\end{equation}
\end{paragrapho}
\medskip

\begin{paragrapho}[Case $n=2$]
The $6$-dimensional space $\mathscr H_{A,1}\otimes \mathscr H_{B,2}$, is made up of the states with one photon in Alice's system, and two in Bob's. It is decomposed into invariant subspaces of dimension 4 and 2,
\begin{equation}
	\mathscr H_{A,1}\otimes \mathscr H_{B,2}=W_{1:2}^{(4)}\oplus W_{1:2}^{(2)}\,
\end{equation}
The subspace $W_{1:2}^{(4)}$ has orthonormal basis
\begin{equation}
\begin{gathered}
\ket{V}_A\ket{(2,0)}_B,\\
\sqrt{\frac{2}{ 3}}\ket{V}_A\ket{(1,1)}_B-\frac{1}{\sqrt 3}\ket{H}_A\ket{(2,0)}_B
,\\
\frac{1}{\sqrt 3}\ket{V}_A\ket{(0,2)}_B
-\sqrt{\frac{2}{3}}\ket{H}_A\ket{(1,1)}_B
,\\
-\ket{H}_A\ket{(0,2)}_B,
\end{gathered}
\end{equation}
and determines the invariant state
\begin{equation}
\rho_{1:2}^{(4)}
=\frac{1}{4}\left(\begin{BMAT}(b)[1pt]{ccc1ccc}{ccc1ccc}
1/3 & 0 & 0 & 0 & -\sqrt{2}/3 & 0 \\
0 & 2/3 & 0 & 0 & 0 & -\sqrt{2}/3 \\
0 & 0 & 1 & 0 & 0 & 0 \\
0 & 0 & 0 & 1 & 0 & 0 \\
-\sqrt{2}/3 & 0 & 0 & 0 & 2/3 & 0 \\
0 & -\sqrt{2}/3 & 0 & 0 & 0 & 1/3 
\end{BMAT}\right).
\end{equation}
The subspace $W_{1:2}^{(2)}$ has orthonormal basis
\begin{equation}
\begin{gathered}
\frac{1}{\sqrt 3}\ket{V}_A\ket{(1,1)}_B+\sqrt{\frac{2}{3}}\ket{H}_A\ket{(2,0)}_B,\\
\sqrt{\frac{2}{3}}\ket{V}_A\ket{(0,2)}_B+\frac{1}{\sqrt 3}\ket{H}_A\ket{(1,1)}_B,
\end{gathered}
\end{equation}
and determines the invariant state
\begin{equation}
\rho_{1:2}^{(2)}=
\frac{1}{2}\left(\begin{BMAT}(b)[1pt]{ccc1ccc}{ccc1ccc}
\mathmakebox[\widthof{$-\sqrt{2}/3$}]{
2/3 }& 0 & 0 & 0 & \sqrt{2}/3 & 0 \\
0 & 1/3 & 0 & 0 & 0 & \sqrt{2}/3 \\
0 & 0 & 0 & 0 & 0 & 0 \\
0 & 0 & 0 & 0 & 0 & 0 \\
\sqrt{2}/3 & 0 & 0 & 0 & 1/3 & 0 \\
0 & \sqrt{2}/3 & 0 & 0 & 0 & 2/3 
\end{BMAT}\right).
\end{equation}
A generic invariant state is a convex combination of $\rho_{1:2}^{(4)}$ and $\rho_{1:2}^{(2)}$, depending on $f \in [0,1]$,
\begin{align}
\rho_{1:2}^\text{(inv)}(f) 
& = f\,\rho_{1:2}^{(2)}+(1-f)\,\rho_{1:2}^{(4)}\nonumber\\
& = \frac{1}{12}
\left(\begin{BMAT}(b)[1pt]{ccc1ccc}{ccc1ccc}
3f+1 & 0 & 0 & 0 & \sqrt{2}(3f-1) & 0 \\
0 & 2 & 0 & 0 & 0 & \sqrt{2}(3f- 1) \\
0 & 0 & 3(1-f) & 0 & 0 & 0 \\
0 & 0 & 0 & 3(1-f) & 0 & 0 \\
\sqrt{2}(3f- 1) & 0 & 0 & 0 & 2 & 0 \\
0 & \sqrt{2}(3f-1) & 0 & 0 & 0 & 3f+1 
\end{BMAT}\right).
\label{inv_12}
\end{align}
\end{paragrapho}
\medskip

\begin{paragrapho}[Case $n=3$]
The $6$-dimensional space $\mathscr H_{A,1}\otimes \mathscr H_{B,3}$, is made up of the states with one photon in Alice's system, and three in Bob's. It is the direct sum of a 5- and a 3-dimensional invariant subspaces, namely,
\begin{equation}
	\mathscr H_{A,1}\otimes \mathscr H_{B,3}=W_{1:3}^{(5)}\oplus W_{1:3}^{(3)}\,.
\end{equation}
The subspace $W_{1:3}^{(5)}$ is spanned by the orthonormal vectors
\begin{equation}
\begin{gathered}
\ket{V}_A\ket{(3,0)}_B,\\
\frac{\sqrt 3}{2}\ket{V}_A\ket{(2,1)}_B
-\frac{1}{2}\ket{H}_A\ket{(3,0)}_B,\\
\frac{1}{\sqrt 2}\bigl(\ket{V}_A\ket{(1,2)}_B
-\ket{H}_A\ket{(2,1)}_B\bigr),\\
\frac{1}{2}\ket{V}_A\ket{(0,3)}_B
-\frac{\sqrt 3}{2}\ket{H}_A\ket{(1,2)}_B,\\
-\ket{H}_A\ket{(0,3)}_B,
\end{gathered}
\end{equation}
and yields the invariant state
\begin{equation}
\rho_{1:3}^{(5)}=
\frac{1}{5}
\left(\begin{BMAT}(b)[1pt]{cccc1cccc}{cccc1cccc}
1/4& 0 & 0 & 0 & 0 & -\sqrt{3}/4 & 0 & 0 \\
0 & 1/2 & 0 & 0 & 0 & 0 & -1/2 & 0 \\
0 & 0 & 3/4 & 0 & 0 & 0 & 0 & -\sqrt{3}/4 \\
0 & 0 & 0 & 1 & 0 & 0 & 0 & 0 \\
0 & 0 & 0 & 0 & 1 & 0 & 0 & 0 \\
-\sqrt{3}/4 & 0 & 0 & 0 & 0 & 3/4 & 0 & 0 \\
0 & -1/2 & 0 & 0 & 0 & 0 & 1/2 & 0 \\
0 & 0 & -\sqrt{3}/4 & 0 & 0 & 0 & 0 & 1/4 
\end{BMAT}\right) .
\end{equation}
The subspace $W_{1:3}^{(3)}$, spanned by the orthonormal vectors
\begin{equation}
\begin{gathered}
\frac{1}{2}\ket{V}_A\ket{(2,1)}_B
+\frac{\sqrt 3}{2}\ket{H}_A\ket{(3,0)}_B,\\
\frac{1}{\sqrt 2}\bigl(\ket{V}_A\ket{(1,2)}_B+\ket{H}_A\ket{(2,1)}_B\bigr),\\
\frac{\sqrt 3}{2}\ket{V}_A\ket{(0,3)}
+\frac{1}{2}\ket{H}_A\ket{(1,2)}_B,
\end{gathered}
\end{equation}
determines the invariant state
\begin{equation}
\rho_{1:3}^{(3)}=
\frac{1}{3}
\left(\begin{BMAT}(b)[1pt]{cccc1cccc}{cccc1cccc}
\mathmakebox[\widthof{$-\sqrt{3}/4$}]{
3/4}& 0 & 0 & 0 & 0 & \sqrt{3}/4 & 0 & 0 \\
0 & 1/2 & 0 & 0 & 0 & 0 & 1/2 & 0 \\
0 & 0 & 1/4 & 0 & 0 & 0 & 0 & \sqrt{3}/4 \\
0 & 0 & 0 & 0 & 0 & 0 & 0 & 0 \\
0 & 0 & 0 & 0 & 0 & 0 & 0 & 0 \\
\sqrt{3}/4 & 0 & 0 & 0 & 0 & 1/4 & 0 & 0 \\
0 & 1/2 & 0 & 0 & 0 & 0 & 1/2 & 0 \\
0 & 0 & \sqrt{3}/4 & 0 & 0 & 0 & 0 & 3/4 
\end{BMAT}\right) .
\end{equation}
A generic invariant state is a convex combination of $\rho_{1:3}^{(5)}$ and $\rho_{1:3}^{(3)}$, depending on $f \in [0,1]$,
\begin{align}
\rho_{1:3}^\text{(inv)}(f)
&=f\,\rho_{1:3}^{(3)}+(1-f)\,\rho_{1:3}^{(5)}\nonumber\\
& = 
\frac{1}{60} 
\scalemath{0.9}{\left(\begin{BMAT}(b)[0pt]{cccc1cccc}{cccc1cccc}
12 f + 3 & 0 & 0 & 0 & 0 & \sqrt{3} (8f -3)  & 0 & 0 \\
0 & 4 f + 6 & 0 & 0 & 0 & 0 & 16 f - 6 & 0 \\
0 & 0 & 9 - 4 f & 0 & 0 & 0 & 0 & \sqrt{3} (8f - 3) \\
0 & 0 & 0 & 12 (1-f) & 0 & 0 & 0 & 0 \\
0 & 0 & 0 & 0 & 12 (1-f) & 0 & 0 & 0 \\
\sqrt{3} (8f - 3) & 0 & 0 & 0 & 0 & 9 - 4 f & 0 & 0 \\
0 & 16 f - 6 & 0 & 0 & 0 & 0 & 4 f + 6 & 0 \\
0 & 0 & \sqrt{3} (8f - 3)  & 0 & 0 & 0 & 0 & 12 f + 3 
\end{BMAT}\right) }.
\label{inv_13}
\end{align}
\end{paragrapho}


\section{Feasibility ranges and bounds for the parameters $Y$ and $c$}
\label{App:ranges}

\noindent
In this section we study in more details the properties of the parameters $Q$ and $c$. In particular, we characterise their range and values on the invariant states.

Recall the definition of the gain $Q$,
\begin{align} 
Q := 
\mathrm{Tr} [
( R_1^H \otimes R_0^V
+ 
R_0^H \otimes R_1^V )
\rho_{B} ] 
\, .
\label{P0def2}
\end{align}
Therefore, the values of $Q$ depends on the eigenvalues of the operator $R_1^H \otimes R_0^V
+ R_0^H \otimes R_1^V$. As shown in Eqs.~(\ref{R0def})-(\ref{R1def}), this operator is diagonal in the number basis,
\begin{align}
R_1^H \otimes R_0^V
+ 
R_0^H \otimes R_1^V
& = 
\sum_{a,b=0}^\infty 
\left[ \lambda_a (1-\lambda_b) 
+
(1-\lambda_a) \lambda_b
\right]
        | a \rangle_H \langle a |
        \otimes
        | b \rangle_V \langle b | \\
& = 
\sum_{a,b=0}^\infty 
\left( \lambda_a + \lambda_b - 2 \lambda_a\lambda_b
\right)
        | a \rangle_H \langle a |
        \otimes
        | b \rangle_V \langle b |  \, .
\end{align}
Recall that eigenvalues depends on the threshold value $\tau$. To make our analysis more concrete we assume $\tau=1$, which is close to the optimal value that maximises the asymptotic key rate. By inspection of the eigenvalues $\lambda_a + \lambda_b - 2 \lambda_a \lambda_b$, we find that the smallest eigenvalue is zero, and is obtained when both $a= b \to \infty$. The largest eigenvalues is $1-\lambda_0$ obtained when $a=0$ and $b \to \infty$. This implies the following range for the gain:
\begin{align}
   Q \in \left( 0 , 1-\lambda_0 \right) \, .
\end{align}

Recall that $Q_j = P_j Y_j$. We now consider the value of $Y_j$ on the invariant state $\rho_{1:j}^\text{(inv)}$, with $j$ photons on Bob side. Since the invariant states commute with the photon number, the reduced state on Bob side is proportional to the projector $\mathbb{P}_j$ into the subspace with $j$ photons:
\begin{align}
    \mathrm{Tr}_A 
    ( \rho_{1:j}^\text{(inv)} )
    = \frac{ \mathbb{P}_j }{\mathrm{Tr} ( \mathbb{P}_j ) }  \, ,
\end{align}
where 
\begin{align}
\mathbb{P}_j =
\sum_{a=0}^j
| a \rangle_H \langle a| \otimes 
| j-a \rangle_V \langle  j-a| \, ,
\end{align}
and
\begin{align}
 \mathrm{Tr}(  \mathbb{P}_j ) = j+1 \, .   
\end{align}

We thus have
\begin{align} 
Y_j = \frac{1}{j+1} \sum_{a=0}^j 
\lambda_{a} + \lambda_{j-a} - 2 \lambda_a \lambda_{j-a} \, .
\end{align}

The behaviour of $Y_{j}$ as a function of $j$ is determined by the threshold value $\tau$. If $\tau$ is sufficiently small, it is a decreasing function of $j$. For example, if we put $\tau=1$, $Y_{j}$ is decreasing for any $j \geq 1$. If instead we put $\tau=2$, $Y_{j}$ is decreasing for any $j \geq 4$. This means that with a careful choice of $\tau$ (not too large) and $k$ (not too small), we can bound the gain $Q$ as follows,
\begin{align}
\sum_{j=0}^k P_j
Y_{j} 
\leq Q
\leq
\sum_{j=0}^k P_j
Y_{j} 
+ 
\left(
1 - \sum_{j=0}^k P_j
\right)
Y_{k+1} 
\, .
\end{align}

Consider now the parameter $c$, defined in Eq.~(\ref{c_def}) as
\begin{align}\label{c_def2}
    c := 
    \frac{1}{2} \mathrm{Tr} \left[ 
    ( 
    |H\rangle \langle H | \otimes  
R_0^H \otimes R_1^V
    +
    |V\rangle \langle V | \otimes 
    R_1^H \otimes R_0^V
    )
    \rho_{AB} 
    \right] 
    \, .
\end{align}
If we restrict to invariant states, we have
\begin{align}\label{c_def2}
    c = 
    \mathrm{Tr} \left[ 
    ( 
    |H\rangle \langle H | \otimes  
R_0^H \otimes R_1^V
    )
    \rho_{AB}^\text{(inv)} 
    \right] 
    \, .
\end{align}
Using the fact that $\mathrm{Tr}_B \rho_{AB} = I/2$ this further simplifies to
\begin{align}\label{c_def2}
    c = \frac{1}{2}
    \mathrm{Tr} \left[ 
    (   
R_0^H \otimes R_1^V
    )
    \rho_{B}^\text{(inv)}(H) 
    \right] 
    \, ,
\end{align}
where $\rho_{B}^\text{(inv)} (H) = \langle H | \rho_{AB}^\text{(inv)} | H \rangle$.

To bound the range of feasibility of the parameter $c$ on invariant state we shall look at the eigenvalues of the operator $R_0^H \otimes R_1^V$. We have
\begin{align}
    R_0^H \otimes R_1^V = 
    \sum_{a,b=0}^\infty (1-\lambda_a) \lambda_b
    | a \rangle_H \langle a | \otimes
    | b \rangle_V \langle b | \, .
\end{align}
As above, we assume $\tau=1$.
The smallest eigenvalues is zero and obtained in the limit $a \to \infty$. The largest eigenvalues is $1-\lambda_0$ and is obtained in the limit of $b \to \infty$. In conclusion, this yields the following feasibility interval for the parameter $c$ on invariant states:
\begin{align}\label{cinter}
    c \in \left( 0, \frac{1-\lambda_0}{2} \right) \, .
\end{align}

Similarly, if consider the invariant state $\rho_{1:j}^\text{(inv)}$ with $j$ photons on Bob side, we obtain the following bounds on the attainable values of $c_{1:j}$:
\begin{align}
\frac{1}{2} (1-\lambda_j) \lambda_0 
\leq c_{j}(f_j) \leq
\frac{1}{2} (1-\lambda_0) \lambda_j \, .
\end{align}
We observe that with increasing $j$, the lower bound becomes smaller and the upper bound becomes larger, yielding the interval (\ref{cinter}) in the limit of $j \to \infty$.

This observation allows us to bound the value of $c$ using a finite number of parameters $c_{1:j}$. We have
\begin{align}
    P_0 c_{0} + \sum_{j=1}^k P_j c_{j}(f_j)
    \leq c \leq
    P_0 c_{0} + \sum_{j=1}^k P_j c_{j}(f_j)
    + 
    \left( 1 - \sum_{j=1}^k P_j \right) \frac{1-\lambda_0}{2} \, .
\end{align}


\section{Properties of the invariant states}
\label{App:inv_states}

\noindent
In this section we present in details the invariant states $\rho_{1:j}^\text{(inv)}$ such that there is one photon on Alice side, and $j$ photons on Bob side. We compute the coefficients $Y_{j}$, $c_{j}$ and the relative entropy $D[\rho_{1:j}^\text{(inv)}]$.

\subsection{Vacuum sector: the invariant state $\rho_{1:0}^\mathrm{(inv)}$}

\noindent
The simplest state corresponds to that with one photon on Alice's side and zero photons on Bob's side. The unique invariant state is
\begin{align}
    \rho_{1:0}^\text{(inv)} = \frac{I}{2} \otimes |0\rangle \langle 0|
    =
    \frac{|H\rangle \langle H| + |V\rangle \langle V|}{2} \otimes |0\rangle \langle 0| \, .
\label{eqn:zero_inv_state}
\end{align}

For this subspace we have
\begin{align}
    Y_0 & = 2 \lambda_0 (1-\lambda_0) \, , \\
    c_{0} & = \frac{1}{2} \lambda_0 (1-\lambda_0) \, ,
\end{align}
and the QBER is, as one would have expected, $E_0 = 2c_0/Y_0 = 1/2$.

The relative entropy for the vacuum sector is
\begin{align}
D[\rho_{1:0}] = 2\lambda_0 (1-\lambda_0) \, .
\end{align}

\subsection{One-photon sector: the invariant states $\rho_{1:1}^\mathrm{(inv)}(f_1)$}

\noindent
Using the expression for the invariant state in Eq.~(\ref{inv_11}) we compute
\begin{align}
    Y_{1} & = \lambda_0 + \lambda_1 - 2\lambda_0 \lambda_1  \, , \\
    c_{1} & = \Tr{( \ket{H} \bra{ H } \otimes R_0^H \otimes R_1^V ) \rho_{1:1}^\text{(inv)} } \\
        & = \frac{ 2f_1 + 1 }{6} \Tr{( R_0^H \otimes R_1^V ) \ket{H} \bra{ H }} 
        + \frac{ 1 - f_1 }{3} \Tr{( R_0^H \otimes R_1^V ) \ket{V} \bra{ V }} \\
        & = \frac{ 2f_1 + 1 }{6} (1-\lambda_1) \lambda_0 
        + \frac{ 1 - f_1 }{3} (1-\lambda_0) \lambda_1 \, .
\end{align}

Defining 
\begin{align}
A_1 := \frac{2f_1+1}{6}, \quad 
B_1 := \frac{4f_1-1}{6}, \quad 
C_1 := \frac{1-f_1}{3}, 
\end{align}
we compute the entropic quantities
\begin{align}
\mathrm{Tr} [ \mathcal{G}( \rho_{1:1}^\text{(inv)} ) \log{ \mathcal{G}(\rho_{1:1}^\text{(inv)} )} ] 
& = (A_1-B_1)(\lambda_0+\lambda_1-2\lambda_0\lambda_1) \log{\left[ (A_1-B_1)(\lambda_0+\lambda_1-2\lambda_0\lambda_1) \right]} \nonumber \\
& \phantom{=}
+ (A_1+B_1)(\lambda_0+\lambda_1-2\lambda_0\lambda_1) \log{\left[ (A_1+B_1)(\lambda_0+\lambda_1-2\lambda_0\lambda_1) \right]} 
\nonumber \\
& \phantom{=}
+ 2 C_1 (\lambda_0+\lambda_1-2\lambda_0\lambda_1) \log{
\left[ C_1 (\lambda_0+\lambda_1-2\lambda_0\lambda_1)
\right]
} \\
& = 
(A_1-B_1)(\lambda_0+\lambda_1-2\lambda_0\lambda_1) \log{(A_1-B_1)} \nonumber \\
& \phantom{=}
+ (A_1+B_1)(\lambda_0+\lambda_1-2\lambda_0\lambda_1) \log{(A_1+B_1)} 
\nonumber \\
& \phantom{=}
+ 2 (A_1+C_1) (\lambda_0+\lambda_1-2\lambda_0\lambda_1) \log{ (\lambda_0+\lambda_1-2\lambda_0\lambda_1) } \nonumber \\
& \phantom{=}
+ 2 C_1 (\lambda_0+\lambda_1-2\lambda_0\lambda_1) \log{C_1}
\end{align}
and
\begin{align}
\mathrm{Tr} [ 
\mathcal{Z}(\mathcal{G}(\rho_{1:1}^\text{(inv)} ))
\log{ \mathcal{Z}(\mathcal{G}(\rho_{1:1}^\text{(inv)} ))}
] 
& = F_- \log{\frac{F_-}{2}}
+ F_+ \log{\frac{F_+}{2}}
\nonumber\\
& \phantom{=}~ + 2 C_1 \lambda_0 (1-\lambda_1) \log{\left[ C_1 \lambda_0 (1-\lambda_1) \right]} 
+ 2 C_1 \lambda_1 (1-\lambda_0) \log{\left[ C_1 \lambda_1 (1-\lambda_0) \right]}
\nonumber \\
& = F_- \log{F_-}
+ F_+ \log{F_+} \nonumber\\
& \phantom{=}~ - 2 A_1 (\lambda_0+\lambda_1-2\lambda_0\lambda_1) \nonumber\\
& \phantom{=}~ + 2 C_1 \lambda_0 (1-\lambda_1) \log{\left[ \lambda_0 (1-\lambda_1) \right]} 
+ 2 C_1 \lambda_1 (1-\lambda_0) \log{\left[ \lambda_1 (1-\lambda_0) \right]} 
\nonumber\\
& \phantom{=}~ + 2 C_1 (\lambda_0 + \lambda_1 - 2\lambda_0\lambda_1) \log{C_1} \, ,
\end{align}
where
\begin{align}
    F_\pm =  
    A_1 (\lambda_0+\lambda_1-2\lambda_0\lambda_1)
    \pm \sqrt{ A_1^2 (\lambda_0-\lambda_1)^2 + 4 B_1^2 \lambda_0 \lambda_1 (1-\lambda_0) (1-\lambda_1)}
    \, .
\end{align}

From them we finally obtain the relative entropy for the single-photon sector:
\begin{align}
D_{1;1}
& = 
- F_+ \log{F_+} 
- F_- \log{F_-}
\nonumber\\
& \phantom{=}~ - 2 C_1 \lambda_0 (1-\lambda_1) \log{\left[ \lambda_0 (1-\lambda_1) \right]} 
- 2 C_1 \lambda_1 (1-\lambda_0) \log{\left[ \lambda_1 (1-\lambda_0) \right]} 
\nonumber\\
& \phantom{=}~+ (A_1-B_1)(\lambda_0+\lambda_1-2\lambda_0\lambda_1) \log{(A_1-B_1)} \nonumber \\
& \phantom{=}~+ (A_1+B_1)(\lambda_0+\lambda_1-2\lambda_0\lambda_1) \log{(A_1+B_1)} 
\nonumber \\
& \phantom{=}~+ 2 A_1 (\lambda_0+\lambda_1-2\lambda_0\lambda_1) \nonumber \\
& \phantom{=}~+ 2 (A_1+C_1) (\lambda_0+\lambda_1-2\lambda_0\lambda_1) \log{ (\lambda_0+\lambda_1-2\lambda_0\lambda_1) } 
\, .
\end{align}


\subsection{Two-photon sector: the invariant states $\rho_{1:2}^\mathrm{(inv)} (f_2)$}

\noindent
Consider the explicit form for the invariant state in Eq.~(\ref{inv_12}). Let us put
\begin{align}
A_2 := \frac{3f_2+1}{12} \, , \, \,
B_2 := \sqrt{2}\frac{3f_2-1}{12} \, , \, \,
C_2 := \frac{1}{6} \, , \, \, 
D_2 := \frac{1-f_2}{4} \, .
\end{align}

The first term in the expression of relative entropy is
\begin{align}
\mathrm{Tr}[ \mathcal{G}(\rho_{1:2}^\text{(inv)} ) \log{\mathcal{G}(\rho_{1:2}^\text{(inv)}} ) ]
& = 2 D_2 (\lambda_0 + \lambda_2 - 2 \lambda_0 \lambda_2) \log{\left[ D_2 (\lambda_0 + \lambda_2 - 2 \lambda_0 \lambda_2) \right]}
+ G_+ \log{\frac{G_+}{2}}
+ G_- \log{\frac{G_-}{2}}
\end{align}
where
\begin{align}
G_\pm 
& = A_2 (\lambda_0 + \lambda_2 - 2\lambda_0\lambda_2) 
+ 2 C_2 \lambda_1 (1-\lambda_1) \nonumber \\
& \phantom{=}~\pm \sqrt{
[ A_2 (\lambda_0 + \lambda_2 - 2\lambda_0\lambda_2) - 2 C_2 \lambda_1 (1-\lambda_1) ]^2
+
8 B_2^2
\lambda_1 (1-\lambda_1)
(\lambda_0 + \lambda_2 - 2\lambda_0\lambda_2)
}
\end{align}

The second term in the expression of relative entropy is
\begin{align}
\mathrm{Tr}[ \mathcal{Z}(\mathcal{G}(\rho_{1:2}^\text{(inv)} )) \log{
\mathcal{Z}(\mathcal{G}(\rho_{1:2}^\text{(inv)} ))}
]
& = 2 D_2 \lambda_0 (1-\lambda_2) \log{\left[ D_2 \lambda_0 (1-\lambda_2)\right]}
+ 2 D_2 \lambda_2 (1-\lambda_0) \log{\left[ D_2 \lambda_2 (1-\lambda_0)\right]} \nonumber \\
& \phantom{=}~ + H_+ \log{\frac{H_+}{2}} + H_- \log{\frac{H_-}{2}}
+ I_+ \log{\frac{I_+}{2}} + I_- \log{\frac{I_-}{2}} \, ,
\end{align}
where
\begin{align}
    H_\pm & = A_2 (1-\lambda_0) \lambda_2 + C_2 \lambda_1 (1-\lambda_1) \pm
    \sqrt{ \left[ A_2(1-\lambda_0)\lambda_2 - C_2 \lambda_1 (1-\lambda_1) \right]^2 + 4 B_2^2 (1-\lambda_0) \lambda_1 (1-\lambda_1) \lambda_2 } \, ,\\
    I_\pm & = A_2 (1-\lambda_2) \lambda_0 + C_2 \lambda_1 (1-\lambda_1) \pm
    \sqrt{ \left[ A_2 (1-\lambda_2) \lambda_0 - C_2 \lambda_1 (1-\lambda_1) \right]^2 + 4 B_2^2 (1-\lambda_2) \lambda_1 (1-\lambda_1)  \lambda_0 } \, .
\end{align}
We then have 
\begin{align}
\begin{split}
    D[ \rho_{1:2}^\text{(inv)} ] & =  2 D_2 \left[ (\lambda_0+\lambda_2(1-2\lambda_0))\log[\lambda_0+\lambda_2(1-2\lambda_0)] - \lambda_0 (1-\lambda_2) \log[\lambda_0(1-\lambda_2)] -\lambda_2 (1-\lambda_0) \log[\lambda_2(1-\lambda_0)]\right] \\
    & \phantom{=}~+ G_+ \log{G_+} + G_- \log{G_-} - H_+ \log{H_+} - H_- \log{H_-} - I_+ \log{I_+} - I_- \log{I_-}\, .
\end{split}
\end{align}

For the two-photon sector we obtain
\begin{align}
    Y_{2} & = \mathrm{Tr}[\mathcal{G}(\rho_{1:2}^\text{(inv)} )] = \frac{2}{3} \left(\lambda_0 + \lambda_1 + \lambda_2 - 2 \lambda_0 \lambda_2 - \lambda_1^2 \right)  \, , \\
    c_{2} & = \Tr{( \ket{H} \bra{ H } \otimes R_0^H \otimes R_1^V ) \rho_{1:2}^\text{(inv)} } = A_2 \lambda_0 (1 - \lambda_2) + C_2 (1 - \lambda_1) \lambda_1 + D_2 (1 - \lambda_0) \lambda_2  \, . 
\end{align}


\subsection{Three-photon sector: the invariant states $\rho_{1:3}^\mathrm{(inv)}(f_3)$}

\noindent
Consider the invariant state in Eq.~(\ref{inv_13}). Defining
\begin{align}
A_3 := \frac{12 f_3 + 3}{60} \, , \, \,
B_3 := \frac{\sqrt{3} (8f_3 - 3)}{60} \, , \, \,
C_3 := \frac{4 f_3 + 6}{60} \, , \, \, 
D_3 := \frac{16 f_3 - 6}{60} \, , \, \, 
E_3 := \frac{9 - 4 f_3}{60} \, , \, \, 
F_3 := \frac{12 (1 - f_3)}{60}\, .
\end{align}
(note that $D_3=2B_3/\sqrt{3}$ and $F_3=3C_3-2A_3$)
such a state reads
\begin{align}
\rho_{1:3}^\text{(inv)}
& = 
\left(\begin{array}{cccc|cccc}
A_3 & 0 & 0 & 0 & 0 & B_3 & 0 & 0 \\
0 & C_3 & 0 & 0 & 0 & 0 & D_3 & 0 \\
0 & 0 & E_3 & 0 & 0 & 0 & 0 & B_3 \\
0 & 0 & 0 & F_3 & 0 & 0 & 0 & 0   \\
\hline
0 & 0 & 0 & 0 & F_3 & 0 & 0 & 0   \\
B_3 & 0 & 0 & 0 & 0 & E_3 & 0 & 0 \\
0 & D_3 & 0 & 0 & 0 & 0 & C_3 & 0 \\
0 & 0 & B_3 & 0 & 0 & 0 & 0 & A_3 
\end{array}\right) \, .
\end{align}

From this expression we compute the relative entropy,
\begin{align}
\begin{split}
D[ \rho_{1:3}^\text{(inv)} ] & = 
(\lambda_1 (1 - 2\lambda_2)+\lambda_2) \left[(C_3 - D_3) \log (C_3 - D_3)+(C_3 + D_3) \log (C_3 + D_3) + 2 C_3 (1 + \log (\lambda_1 (1 - 2\lambda_2)+\lambda_2))\right] \\
& + 2 F_3\left[(\lambda_0+\lambda_3-2 \lambda_0 \lambda_3) \log (\lambda_0+\lambda_3-2 \lambda_0 \lambda_3) - \lambda_0 (1-\lambda_3) \log (\lambda_0 (1-\lambda_3)) - \lambda_3 (1-\lambda_0) \log (\lambda_3 (1-\lambda_0))\right] \\
& + P_-\log P_- + P_+\log P_+ - Q_-\log Q_- - Q_+\log Q_+ - R_- \log R_- - R_+\log R_+ - S_- \log S_- - S_+\log S_+ 
\, ,
\end{split}
\end{align}
where
\begin{align}
\begin{split}
P_\pm & = A_3 (\lambda_0+\lambda_3-2\lambda_0\lambda_3)
+ E_3 (\lambda_1+\lambda_2(1-2\lambda_1)) \\
& \phantom{=}~\pm
\sqrt{ 
\left[ A_3 (\lambda_0+\lambda_3(1-2\lambda_0)) - E_3 (\lambda_1+\lambda_2(1-2\lambda_1)) \right]^2
+ 4 B_3^2 (\lambda_3+\lambda_0(1-2\lambda_3)) (\lambda_2+\lambda_1(1-2\lambda_2)) } \, , \\
Q_\pm 
    & = C_3 (\lambda_1 + \lambda_2 (1-2\lambda_1)) \pm 
    \sqrt{ C_3^2 (\lambda_1 - \lambda_2)^2 + 4 D_3^2 \lambda_1 \lambda_2 (1 - \lambda_1) (1 - \lambda_2)} \,, \\
R_\pm  
    & = A_3 \lambda_0 (1-\lambda_3) 
    + E_3 \lambda_1 (1-\lambda_2)  
    \pm 
    \sqrt{ 
    \left[ A_3 \lambda_0 (1-\lambda_3) - E_3 \lambda_1 (1-\lambda_2) \right]^2
    + 4 B^2 \lambda_0 \lambda_1 (1-\lambda_2) (1-\lambda_3) } \, ,\\
S_\pm  
    & = A_3 (1-\lambda_0) \lambda_3 
    + E_3 (1-\lambda_1) \lambda_2  
    \pm 
    \sqrt{ 
    \left[ A_3 (1-\lambda_0) \lambda_3 - E_3 (1-\lambda_1) \lambda_2 \right]^2
    + 4 B^2 (1-\lambda_0) (1-\lambda_1) \lambda_2 \lambda_3 } \, .
\end{split}
\end{align}

Finally, for the three-photon sector we obtain
\begin{align}
    Y_{3} & = \mathrm{Tr}[\mathcal{G}(\rho_{1:3}^\text{(inv)} )] = \frac{1}{2} \left(\lambda_0 + \lambda_1 + \lambda_2 + \lambda_3 - 2 \lambda_0 \lambda_3 - 2 \lambda_1 \lambda_2\right)  \, , \\
    c_{3} & = \Tr{( \ket{H} \bra{ H } \otimes R_0^H \otimes R_1^V ) \rho_{1:3}^\text{(inv)} } = A_3 \lambda_0 (1 - \lambda_3) + C_3 \lambda_1 (1 - \lambda_2) + E_3 (1 - \lambda_1) \lambda_2 + F_3 (1 - \lambda_0) \lambda_3 \, .
\end{align}


\section{Invariant states under Gaussian noise}\label{App:Gauss}

\noindent
In this Appendix we derive the invariant states for a communication channel characterised by loss $\eta$ and Gaussian noise with variance $N$.

In the EB representation, first the lossy channel is applied to state (\ref{inputstate}), yielding
\begin{align}
    & \frac{\eta}{2} 
    \left( 
    |H\rangle \langle H | \otimes |H\rangle \langle H |
        + |V\rangle \langle V | \otimes |V\rangle \langle V | 
+ |H\rangle \langle V | \otimes |H\rangle \langle V |
    + |V\rangle \langle H | \otimes |V\rangle \langle H |
    \right) + \frac{(1-\eta)}{2} I \otimes |0\rangle \langle 0| \otimes |0\rangle \langle 0| \, .
\end{align}

Second, a Gaussian-noise channel is applied to each mode belonging to Bob. Recall the action on each mode of this noise:
\begin{align}\label{noisech2}
\rho \to \int \frac{d^2\alpha}{\pi N}
e^{-|\alpha|^2/N}
\mathcal{D}(\alpha) \rho \mathcal{D}(\alpha)^\dag 
\, ,
\end{align} 
where $\mathcal{D}(\alpha)$ is the displacement operator.

To compute the effect of this noise, we first apply independent displacements on the H and V modes, obtaining the state
\begin{align}
    &  
    \frac{\eta}{2} 
    \left( 
    |H\rangle \langle H | \otimes \mathcal{D}(\alpha) |H\rangle \langle H | \mathcal{D}(\alpha)^\dag
        + |V\rangle \langle V | \otimes \mathcal{D}(\beta) |V\rangle \langle V | \mathcal{D}(\beta)^\dag \right. \nonumber \\ 
    & \left. + |H\rangle \langle V | \otimes \mathcal{D}(\alpha) |H\rangle \langle V | \mathcal{D}(\beta)^\dag
    + |V\rangle \langle H | \otimes \mathcal{D}(\beta) |V\rangle \langle H | \mathcal{D}(\alpha)^\dag
    \right) \nonumber \\
    & + \frac{(1-\eta)}{2} I \otimes \mathcal{D}(\alpha) |0\rangle \langle 0| \mathcal{D}(\alpha)^\dag
    \otimes \mathcal{D}(\beta) |0\rangle \langle 0| \mathcal{D}(\beta)^\dag \, .
\end{align}

Finally, we will average over the displacement amplitudes $\alpha$ and $\beta$.

The result of the average is immediate for the terms of the kind $\mathcal{D}(\beta) |0\rangle \langle 0| \mathcal{D}(\beta)^\dag$, where the displacement is applied on the vacuum state. The terms yield thermal states with $N$ mean photon number,
\begin{align}
    \rho_N = \frac{1}{N+1} \sum_{k=0}^\infty \left( \frac{N}{N+1} \right)^k
    \, .
\end{align}
Below we use the notation $\rho_N^{(H)}$ and $\rho_N^{(V)}$ to indicate a thermal state in horizontal or vertical mode of polarisation.

To compute the other terms, we exploit the expansion in number basis of the displacement operator:
\begin{align}
    \langle m | \mathcal{D}(\alpha) | n \rangle =
    \sqrt{ \frac{m!}{n!} } \, (-\alpha^*)^{n-m} e^{-|\alpha|^2/2}
    L_m^{(n-m)}(|\alpha|^2)
        \, \, \, \, 
    \text{for} \, \, m < n \, .
\end{align}

The only non-zero terms are:
\begin{align}
\int \frac{d^2\alpha}{\pi N}
e^{-|\alpha|^2/N}
\langle 0 | D(\alpha) |1\rangle \langle 1 | D(\alpha)^\dag |0\rangle
= \frac{N}{(N+1)^2}
\end{align}
and
\begin{align}
\int \frac{d^2\alpha}{\pi N}
e^{-|\alpha|^2/N}
\langle m | D(\alpha) |1\rangle \langle 1 | D(\alpha)^\dag |m\rangle
& = \frac{1}{N+1} \left( \frac{N}{N+1} \right)^{m} \frac{m + N^2}{N(N+1)}  \, .
\end{align}

From this we define a probability distribution
\begin{align}
    p_m := \left\{
\begin{array}{lcr}
\frac{N}{(N+1)^2} & \,\, \text{if} \,\, & m=0 \\
\frac{1}{N+1} \left( \frac{N}{N+1} \right)^{m} \frac{m + N^2}{N(N+1)} & \,\, \text{if} \,\, & m \geq 1 
\end{array}
    \right.
\end{align}

For the off-diagonal terms, the only non-zero contributions come from
\begin{align}
\int \frac{d^2\alpha}{\pi N}
e^{-|\alpha|^2/N}
\langle m+1 | D(\alpha) |1\rangle \langle 0 | D(\alpha)^\dag |n\rangle
& = \frac{1}{N+1} \left( \frac{N}{N+1} \right)^{m} \frac{\sqrt{m+1}}{N+1} \, .
\end{align}

From this we define the coefficients 
\begin{align}
    t_m := \frac{1}{N+1} \left( \frac{N}{N+1} \right)^{m} \frac{\sqrt{m+1}}{N+1} \, .
\end{align}

In conclusion, the state after averaging over the noise realisation is
\begin{align}
    \rho_{AB}^\text{(inv)} & = \frac{\eta}{2} 
    \left( 
    |H\rangle \langle H | \otimes 
    \sum_m p_m |m_H \rangle \langle m_H| 
    \otimes \rho_N^{(V)}
        + |V\rangle \langle V | \otimes 
    \rho_N^{(H)} \otimes
    \sum_m p_m |m_V \rangle \langle m_V|  \right. \nonumber \\ 
    & \phantom{=}~ \left. + |H\rangle \langle V | \otimes 
    \sum_m t_m |m+1\rangle \langle m| 
    \otimes 
    \sum_{m'} t_{m'} |m'\rangle \langle m'+1|
    + \mathrm{h.c.}
    \right) \nonumber \\
    & \phantom{=}~ + (1-\eta) \frac{I}{2} \otimes \rho_N^{(H)} \otimes \rho_N^{(V)} \, .
    \label{invstateTh}
\end{align}


\subsection{Invariant state in the vacuum sector for Gaussian noise}
\label{sec:invar_state}

\noindent
From Eq.~(\ref{invstateTh}) we obtain the (not-normalised) invariant state with one photon on Alice side and the vacuum on Bob side:
\begin{align}
    P_0 \rho_{1:0}^\text{(inv)}
    & = \frac{I}{2} \otimes \left( \eta \frac{p_0}{N+1} + (1-\eta)  
    \left( \frac{1}{N+1} \right)^2 
    \right)
    |0 \rangle \langle 0|
    \, .
\end{align}

By computing the trace we obtain 
\begin{align}
    P_0 = 
    \eta \frac{p_0}{N+1} + (1-\eta)  
    \left( \frac{1}{N+1} \right)^2
    \, .
\end{align}

\subsection{Invariant state in the one-photon sector for Gaussian noise}

From Eq.~(\ref{invstateTh}) we obtain the (not-normalised) invariant state with one photon on Alice side and one photon on Bob side.

In the basis $\{ |HH\rangle , |HV\rangle , |VH\rangle , |VV\rangle\}$, it reads
\begin{align}
P_1 \rho_{1:1}^\text{(inv)} 
& \equiv
\frac{\eta}{2}
\left( 
\begin{array}{cc|cc}
 \frac{p_1}{N+1} & 0 & 0 & t_0^2 \\
0 &  \frac{ p_0}{N+1} \frac{N}{N+1} & 0 & 0 \\
\hline
0 & 0 & \frac{p_0}{N+1} \frac{N}{N+1} & 0 \\
 t_0^2 & 0 & 0 & \frac{p_1}{N+1} 
\end{array}
\right) 
+ \frac{1-\eta}{2}  \left( \frac{1}{N+1}\right)^2 \frac{N}{N+1}
\left( 
\begin{array}{cc|cc}
1 & 0 & 0 & 0 \\
0 & 1 & 0 & 0 \\
\hline
0 & 0 & 1 & 0 \\
0 & 0 & 0 & 1 
\end{array}
\right) \, .
\end{align}

The trace gives
\begin{align}
    P_1 = \frac{\eta}{N+1} \left(
     p_0 \frac{N}{N+1} 
     + p_1
     \right)
    + 2(1-\eta) \left( \frac{1}{N+1}\right)^2 \frac{N}{N+1} \, .
\end{align}

\subsection{Invariant state in the two-photon sector for Gaussian noise}

\noindent
From Eq.~(\ref{invstateTh}) we obtain the (not-normalised) invariant state with one photon on Alice side and two photons on Bob side.

In the basis
\begin{align}
\{ \ket{H;(2,0)} , \ket{H;(1,1)} , \ket{H;(0,2)} ,
\ket{V;(2,0)} , \ket{V;(1,1)} , \ket{V;(0,2)}
\} \, ,
\end{align}
the (not-normalised) invariant state has the following matrix representation:
\begin{align}
P_2 \rho_{1:2}^\text{(inv)} 
& \equiv
\frac{\eta}{2}
\left(\begin{array}{ccc|ccc}
\frac{p_2}{N+1} & 0 & 0 & 0 & t_0t_1 & 0 \\
0 & \frac{p_1}{N+1} \frac{N}{N+1} & 0 & 0 & 0 & t_0t_1 \\
0 & 0 & \frac{p_0}{N+1} \left( \frac{N}{N+1} \right)^2 & 0 & 0 & 0 \\
\hline
0 & 0 & 0 & \frac{p_0}{N+1} \left( \frac{N}{N+1} \right)^2 & 0 & 0 \\
t_0t_1 & 0 & 0 & 0 & \frac{p_1}{N+1} \frac{N}{N+1} & 0 \\
0 & t_0t_1 & 0 & 0 & 0 & \frac{p_2}{N+1} 
\end{array}\right)  
+
\frac{1-\eta}{2} \left( \frac{1}{N+1} \frac{N}{N+1}\right)^2
I_6 \, ,
\end{align}
where $I_6$ is the $6 \times 6$ identity matrix.

By computing the trace we obtain
\begin{align}
P_2 = 
\eta \left(
 \frac{p_0}{N+1} \left( \frac{N}{N+1} \right)^2
+ \frac{p_1}{N+1} \frac{N}{N+1}
+\frac{p_2}{N+1}
\right)
+ 3 (1-\eta) 
\left( \frac{1}{N+1} \frac{N}{N+1}\right)^2 \, .
\end{align}

\subsection{Invariant state in the three-photon sector for Gaussian noise}

\noindent
From Eq.~(\ref{invstateTh}) we obtain the (not-normalised) invariant state with one photon on Alice side and three photons on Bob side.

In the basis
\begin{align}
\{
\ket{H;(3,0)} ,
\ket{H;(2,1)} ,
\ket{H;(1,2)} ,
\ket{H;(0,3)} , 
\ket{V;(3,0)} ,
\ket{V;(2,1)} ,
\ket{V;(1,2)} ,
\ket{V;(0,3)}
\} \, ,
\end{align}
we have
\begin{align}
P_3 \rho_{1:3}^\text{(inv)}
& \equiv
\frac{\eta}{2} 
\left(\begin{array}{cccc|cccc}
\frac{p_3}{N+1} & 0 & 0 & 0 & 0 & t_0 t_2 & 0 & 0 \\
0 & \frac{p_2}{N+1} \frac{N}{N+1} & 0 & 0 & 0 & 0 & t_1^2 & 0 \\
0 & 0 & \frac{p_1}{N+1} \left(\frac{N}{N+1}\right)^2 & 0 & 0 & 0 & 0 & t_0 t_2 \\
0 & 0 & 0 & \frac{p_0}{N+1} \left(\frac{N}{N+1}\right)^3 & 0 & 0 & 0 & 0 \\
\hline
0 & 0 & 0 & 0 & \frac{p_0}{N+1} \left(\frac{N}{N+1}\right)^3 & 0 & 0 & 0 \\
t_0 t_2 & 0 & 0 & 0 & 0 & \frac{p_1}{N+1} \left(\frac{N}{N+1}\right)^2 & 0 & 0 \\
0 & t_1^2 & 0 & 0 & 0 & 0 & \frac{p_2}{N+1} \frac{N}{N+1} & 0 \\
0 & 0 & t_0 t_2 & 0 & 0 & 0 & 0 & \frac{p_3}{N+1} 
\end{array}\right)  \nonumber \\
& \phantom{=}~+ \frac{1-\eta}{2} \left( \frac{1}{N+1} \right)^2 \left( \frac{N}{N+1}\right)^3
I_8
 \, ,
\end{align}
where $I_8$ is the $8 \times 8$ identity matrix.

From the trace we obtain
\begin{align}
P_3 =
\eta \left(
\frac{p_0}{N+1} \left(\frac{N}{N+1}\right)^3
+ \frac{p_1}{N+1} \left(\frac{N}{N+1}\right)^2
+ \frac{p_2}{N+1} \frac{N}{N+1}
+ \frac{p_3}{N+1} 
\right)
+ 4 ( 1-\eta) \left( \frac{1}{N+1} \right)^2 \left( \frac{N}{N+1}\right)^3 \, .
\end{align}
This concludes all the properties for each invariant state up to the three-photon subspace.


\end{widetext}

\end{document}